\documentclass[twocolumn]{aastex631}

\newcommand{\fermata}{\includegraphics[width=4.5mm]{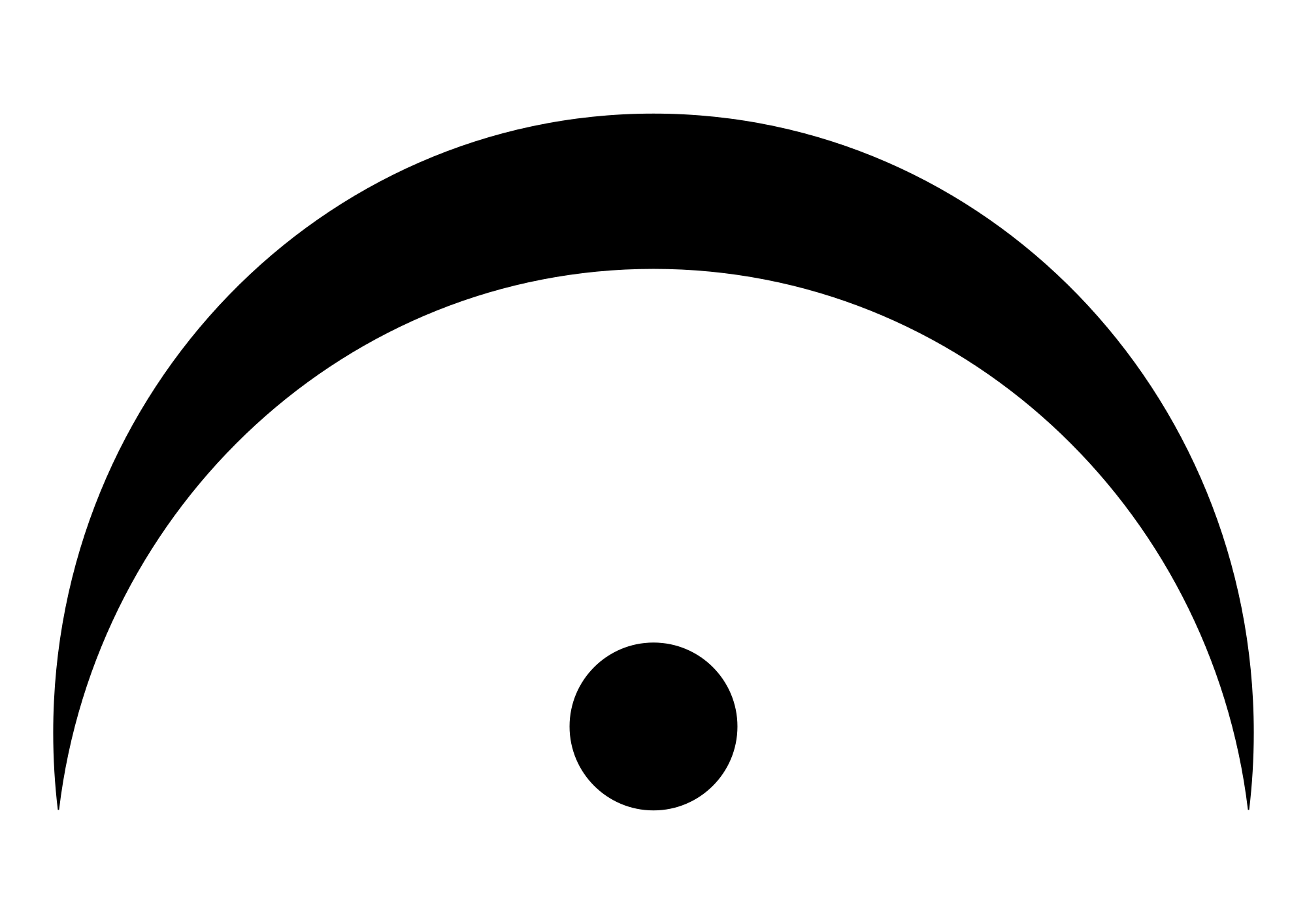}}
\newcommand{\kms}{\,km\,s$^{\rm -1}$\,}
\newcommand{\vg}{$v_{\rm gas}$}
\newcommand{\vd}{$v_{\rm dust}$}
\newcommand{\vdr}{$v_{\rm drift}$}

\usepackage{longtable}

\begin{document}

\title[Dust Around RT~Vir]{ Unraveling the Dusty Environment Around RT~Vir}

\author[0009-0006-0724-7331]{Michael D. Preston}
\affiliation{University of Texas at San Antonio \\
1 UTSA Circle \\
San Antonio, TX 78249, USA}

\author[0000-0001-7832-0344]{Angela K. Speck}
\affiliation{University of Texas at San Antonio \\
1 UTSA Circle \\
San Antonio, TX 78249, USA}

\author[0000-0003-2983-5717]{Sean Dillon}
\affiliation{University of Texas at San Antonio \\
1 UTSA Circle \\
San Antonio, TX 78249, USA}

\author[0000-0001-9855-8261]{Beth Sargent}
\affiliation{Department, Space Telescope Science Institute\\
3700 San Martin Drive\\
Baltimore, MD 21218, USA}
\affiliation{John's Hopkins University\\
3400 N Charles Street\\
Baltimore, MD 21218, USA}

\begin{abstract}

Infrared studies of asymptotic giant branch (AGB) stars are critical to our understanding of the formation of cosmic dust. In this investigation, we explore the mid-to-far-infrared emission of oxygen-rich AGB star RT~Virginis. 
This optically thin dusty environment has unusual spectral features when compared to other stars in its class. 
To explore this enigmatic object we use the 1-D radiative transfer modeling code DUSTY. 
Modeled spectra are compared with observations from the Infrared Space Observatory (ISO), InfraRed Astronomical Satellite (IRAS), the Herschel Space Observatory and a host of other sources to determine the properties of RT~Vir's circumstellar material.  Our models suggest a set of two distant and cool  dust shells  at low optical depths ($\tau_{V, \rm inner} = 0.16$, $\tau_{V, \rm outer} = 0.06$), with inner dust temperatures: $\mathrm{T_1 = 330K, T_3 = 94K}$.  Overall, these dust shells exhibit a chemical composition consistent with dust typically found around O-rich AGB stars. However, the distribution of materials differs significantly. The inner shell consists of a mixture of silicates, Al$_2$O$_3$, FeO, and Fe, while the outer shell primarily contains crystalline Al$_2$O$_3$ polymorphs. This chemical change is indicative of two distinct epochs of dust formation around RT~Vir. These changes in dust composition are driven by  either changes in the pressure-temperature conditions around the star or by a decrease in the C/O ratio due to hot-bottom burning.

\end{abstract}


\section{Introduction} \label{sec:intro}

Understanding the nature and formation of cosmic dust is critical to understanding most cosmic environments. Infrared (IR) Astronomy has shown that dust contributes to the physics of star and planet formation, mass loss from evolved stars, interstellar gas heating, and the formation of molecules \citep[e.g.][]{Draine_2003, Krishna-Swamy_2005, Woitke_2006, Krugel_2008}. 

In our galaxy, some of the most important dust formation regions are around Asymptotic Giant Branch (AGB) stars \citep{Kwok_2004}. AGB stars evolve from low-to-intermediate mass main-sequence stars ( 0.8--8M$_\odot$). Pulsations and shocks propagating through the upper atmospheres of AGB stars lift material off the stellar surface, creating expanding shells of material around the star \citep{Hofner_2016}. The environment in these shells is relatively cool with little ultraviolet (UV) radiation from the star, making it ideal for the formation of molecules and condensation of dust grains. The dust grains are higher in opacity, and thus are more susceptible to radiation pressure from the star. As these dust grains are driven away from the star, they drag the gas along with them. This circumstellar material continues to move away from the star, and can form shells of coupled gas \& dust around the central star \citep{Hofner_2009}.

Dust chemistry in these circumstellar outflows is dictated by the abundance of carbon and oxygen in the outflow. Carbon and oxygen will preferentially form neutral CO gas, sequestering the less abundant element and restricting the chemistry of the molecules and dust grains that form later. Consequently, if C is more abundant than O, the primary dust chemistry will be C-rich, and grains of SiC, graphite (or amorphous C) and other C-bearing minerals will form. Conversely, O-rich stars will form silicate minerals ((SiO$_4$)$^{-4}$ tetrahedra-bearing) and metal-oxides (Al$_2$O$_3$, FeO, etc.)  In systems like these (C rich or O rich), grain condensation is not fully understood \citep[e.g.][]{Woitke_2006, Hofner_2009}. In the interest of exploring dust production in O-rich AGB outflows we investigate the dusty environment around the star RT~Virginis (RT~Vir).

\subsection{RT~Virginis}

\begin{table}
  \centering
     \caption{Table of basic properties of RT~Vir  \label{tab:RTVir}}
     \begin{tabular}{llc}
    \hline\hline
    Property & Value & Reference\\
    \hline
    Spectral Type & M8III & 1\\
    Variability Type & SRb & {2}\\
    Period (days) & 153--320 & 3\\
    Mass-loss rate ($\dot{M}$/\,M$_\odot {\rm yr}^{-1}$) & $10^{-7}-10^{-6}$ & 4, 5\\
    CO expansion velocity (\vg) & 7.8, 11.3\kms & 6, 7 \\
    Luminosity ($L_\star/L_\odot$) & $5012^{+1,154}_{-938}$  & 5 \\
    Stellar temperature (T$_{\rm eff}$, K) & 2902 & 8\\
     Maser emission type    & H$_2$O &  9, 10\\
     Hipparcos Distance & 135$\pm$15\,pc & 11\\
    GAIA/VLTI Distance & 226$\pm$7\,pc & 12, 13\\
    \hline\hline
    \end{tabular}
   \begin{tabular}{p{\columnwidth}}
   References:
    (1) \citet{Joy_1942};
    (2) \citet{Van_Der_Veen_1995};
    (3) \citet{Kudashkina_2022};
    (4) \citet{Imai_2003};
    (5) \citet{Brand_2020};
    (6) \citet{Olofsson_2002}
    (7) \citet{Loup_1993};
    (8) \citet{Sharma_2016}
    (9) \citet{Bains_2003};
    (10) \citet{Kim_2010};
    (11) \citet{Van_Leeuwen_2007};
    (12) \citet{Zhang_2017};
    (13) \citet{Andriantsaralaza_2022}.
    \end{tabular}
\end{table}

\begin{figure}
    \centering
    \includegraphics[width=\columnwidth]{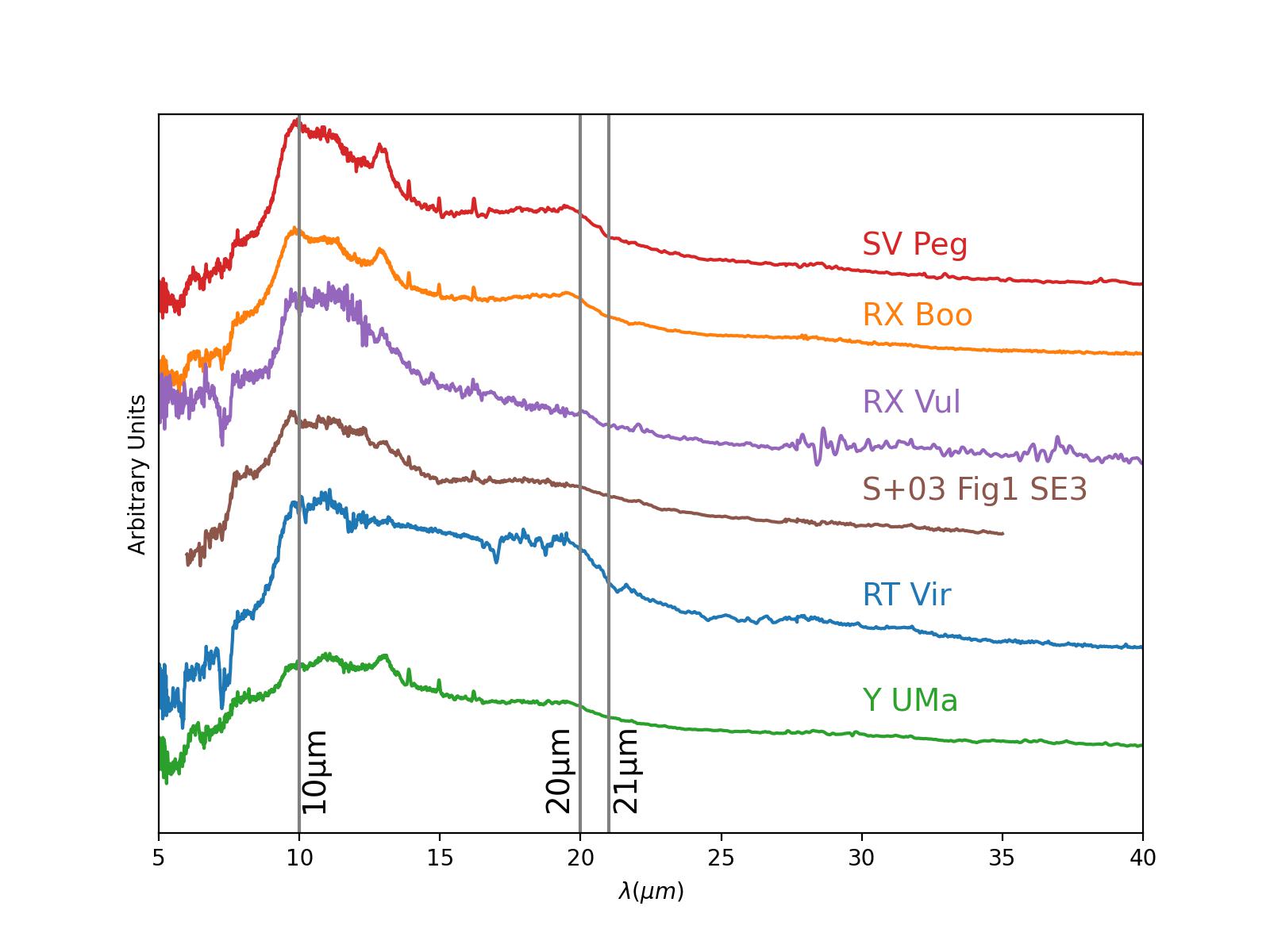}
    \caption{Comparison of mid-infrared photosphere-subtracted ISO-SWS spectra of RT~Vir with similar types of stars,  SV Peg, RX Boo, RX Vul and Y UMa, along with the average SE3 class spectrum (without a 13\,$\mu$m feature) presented by \citet{Sloan_2003}. For the individual stars, the photosphere subtraction uses a model  M5\,III-Star template courtesy of Kevin Volk,  described in \S~\ref{sec:obs}. The ISO spectra used in \citet{Sloan_2003} to generate their SE class averages where also stellar subtracted, but using ISO data for NU~Pav, a naked M6\,III-star and the photosphere to be subtracted}
    \label{fig:compare}
\end{figure}

RT~Vir is an AGB star first described by Williamina Fleming \citep{Pickering_1896}, and its vital statistics are given in Table~\ref{tab:RTVir}.
RT~Vir is a target in many studies of astrophysical maser emission and in several surveys of AGB and semiregular (SR) variable stars. The circumstellar emission has been studied extensively through {\em imaging} in the IR \citep[via the Herschel \& AKARI Space Observatories;][]{Pilbratt_2010, Kawada_2007} and the ultraviolet \citep[UV; via the Galaxy Evolution Explorer, GALEX;][]{Martin_2005} 
\citep[e.g.,][]{Groenewegen_2011,Mayer_2011,Cox_2012,Sacuto_2013,Paladini_2017, Sahai_2023}.
These studies lend a great deal of insight into the properties of AGB stars in general and specifically RT~Vir. However, there is a dearth of investigations into this star's circumstellar dust spectroscopy in the mid-to-far-IR.  Observations of RT~Vir at mid-IR wavelengths show unusual emission features between 9-40\,$\mu$m as shown in Figure~\ref{fig:compare}. To make a fair comparison to other AGB star spectra, we selected only those objects classified as
SE3t  in \citet{Sloan_Price_1998}, with spectral types later than M6.5III and variability type SRb. However, since only two such objects met these criteria and had available ISO SWS spectra, we also included two additional stars: SE2t star, Y UMa, and SE3 star RX~Vul. RX~Vul is a Mira variable rather than a SR type, but the SE3 class (without the 13\,$\mu$m feature) only includes 4 stars so the choices were sparse. We also included data from Figure 1 of \cite{Sloan_2003} that depicts an averaged spectrum of stars with SE3 dust emission, and little to no emission at 13\,$\mu$m. 
AGB stars exhibit a wide range of spectral features \citep[e.g.,][]{Speck_2000,Little-Marenin_Little_1988,Sloan_2003}, so these measures were taken to minimize spectral differences between the selected stars and RT~Vir.

In Figure~\ref{fig:compare}, for RX~Boo, SV~Peg and Y~Uma we see significant IR enhancement at 
$\sim$10 and 18\,$\mu$m due to warm silicates, and some other features at $\sim$13, 20, 21\,$\mu$m correlated with other oxides present in the circumstellar environment.  RX Vul shows a strong enhancement at $\sim$10\,$\mu$m but lacks the emission features at 13, 20 and 21\,$\mu$m associated with oxides.
While the spectrum of RT~Vir does exhibit features in the mid- and far-IR, they are unlike those seen in the comparison O-rich AGB stars. Figure~\ref{fig:compare}  also shows that RT~Vir lacks a pronounced 13\,$\mu$m feature, instead showing a ``bridge'' of emission between features at 10 and 20\,$\mu$m.  Additionally, this ``bridge'' emission distinguishes RT~Vir from SE3 class stars like RX~Vul which also lacks a 13\,$\mu$m feature. This demonstrates oddity, even allowing for the intrinsic variation within this class of objects.
This deviation from the classical O-rich AGB dust shell features makes RT~Vir an even more interesting object to study.

Here we present new radiative transfer models of this unusual AGB star. In \S~\ref{sec:obs} we describe from whence the observation data were collected. 
In \S~\ref{sec:modeling}, we describe the methodology used to produce radiative transfer models that fit well to the observations.
In \S~\ref{sec:results} we provide the results of the modeling, and \S~\ref{sec:disc} discusses the implications and interpretation of the modeling results in terms of dust composition, the large extent of the dust shell, and mass-loss timescales.
Conclusions are drawn in \S~\ref{sec:conc}.

\section{Observations} \label{sec:obs}

We have obtained photometric observations from UV to IR wavelengths for RT~Vir from multiple sources including Simbad\citep{Ducati_2002} and IRSA. We also collected fully processed post-pipeline spectral data from the Infrared Space Observatory's Short Wavelength Spectrometer \citep[ISO SWS;][]{Kessler_1996,de_Graauw_1996}, which were acquired from an online atlas\footnote{Detailed data reduction information is available from the atlas website:   \url{https://users.physics.unc.edu∼gcsloan/library/swsatlas/aot1.html.}} associated with \cite{Sloan_2003}.  The resulting spectral energy distribution (SED) of RT~Vir 
is shown in Figure~\ref{fig:SED} , while the photometric observations are listed in Table~\ref{tab:fullPhot} in the Appendix.
The vertical spread in photometry points is attributable to the intrinsic variability of RT~Vir.
Also included in Figure~\ref{fig:SED} are labels for the locations of prominent molecular absorption features in the near- to mid-IR, the precise wavelengths for which are listed in Table~\ref{tab:molecules}. The molecules that cause these absorption features occur both in the stellar photosphere and in the circumstellar outflow. For this reason, we have included in the figure a template stellar spectrum  (blue line) which we will use in radiative transfer modeling (see \S~\ref{sec:modeling}). 
Our stellar spectrum template (courtesy of Kevin Volk) is compiled from a collection of observations. In the infrared ISO SWS and LWS data for several M5III stars were compiled, giving high-resolution data from 2--200\,$\mu$m. The visible observations were sourced from \citet{Pickles1985}.
The discrepancy between the template and the observed SED at $\lambda > 8\,\mu$m arises from the dust excess, which we model in this study. At $\lambda < 1,\mu$m, the poor fit is primarily due to a mismatch in spectral type: RT~Vir is an M8III star, whereas the template spectra correspond to M5III stars. The higher temperature of M5 \citep[$T_{\rm eff}\sim$3400\,K,][]{1998A&A...331..619P} stars compared to M8 ($T_{\rm eff}\sim$2800\,K) stars significantly affects the short (visible/UV) wavelengths, resulting in the lower observed fluxes at visible wavelengths. In contrast, the mid-IR differences are much less pronounced. Since our modeling primarily focuses on IR wavelengths, achieving a precise fit at visible and UV wavelengths is not critical. It is also worth noting that it is almost impossible to find a reference for a reliable list of wavelengths and line widths simply because of the complex nature of molecular spectra. This problem is exacerbated by the fact that the CO, H$_2$O, and SiO absorption features in the 3-10\,$\mu$m range are broad and overlap, sometime merging into a single broader absorption region \citep[e.g.,][]{Gustafsson_1998}, as seems to be the case for RT~Vir. The references quoted in Table~\ref{tab:molecules} provide some wavelengths or ranges for specific molecular species.

\begin{figure}
        \includegraphics[width=\columnwidth]
    {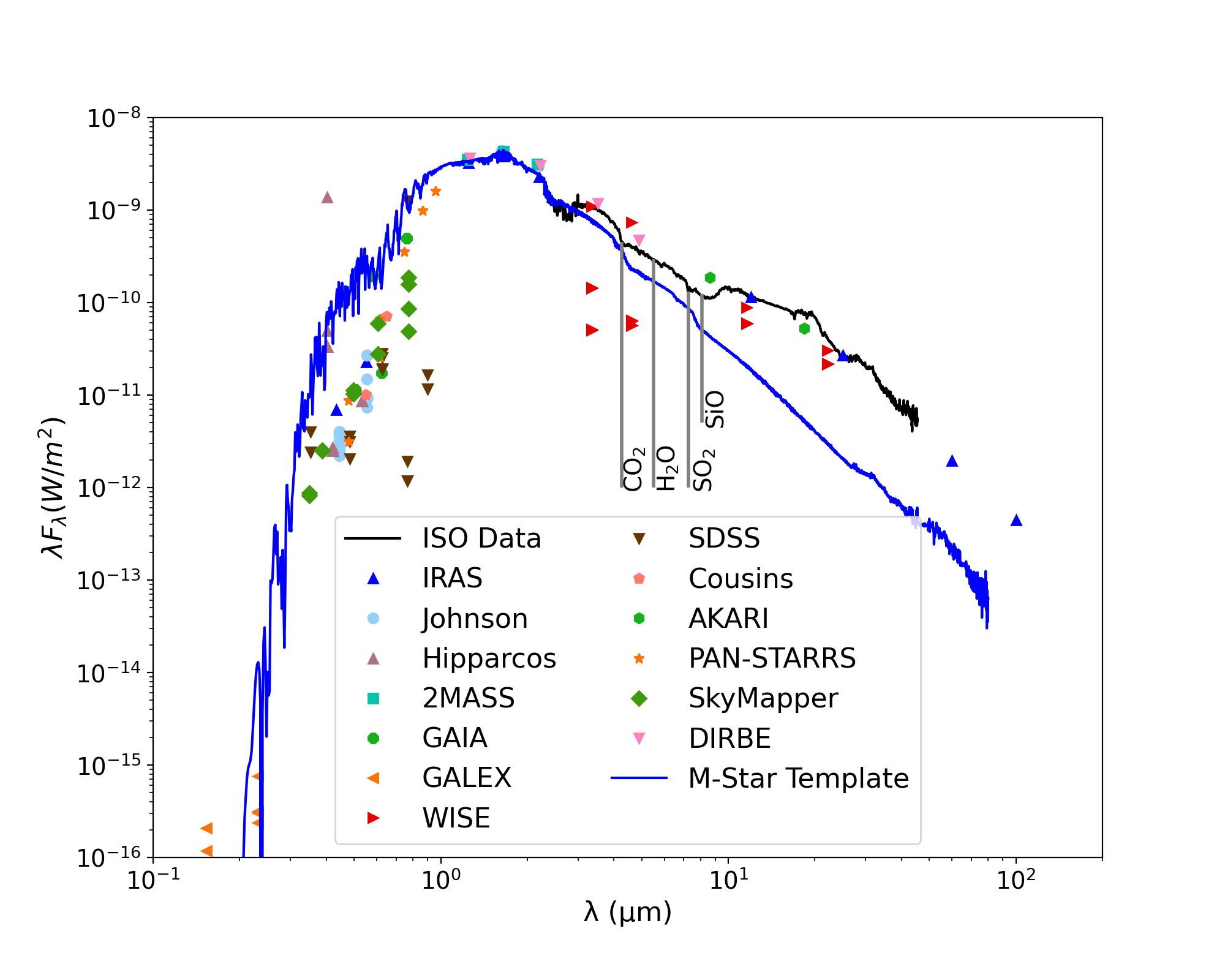}
    \caption{Spectral Energy Distribution (SED) for RT~Vir. The dark line is spectroscopy from ISO SWS, and the solid blue line is the  M~Star  template described in the text. All other marks are photometry measurements from various surveys performed at different times. The large spread in brightness at a given wavelength are a result of the variable nature of RT~Vir. Grey vertical lines indicate positions of molecular absorption bands in RT Vir's atmosphere} 
    \label{fig:SED}
\end{figure}

\begin{table}
    \centering
    \caption{Wavelengths of mid-infrared molecular absorption features}
     \begin{tabular}{lr}
    \hline\hline
    Molecule   &  Wavelengths ($\mu$m)\\
    \hline     
    CO     &  1.4, 2.3, 4.6\\
    H$_2$O &  1.5, 1.9, 2.7, 6.2 \\
    SiO    &  8.2 \\ 
    SO$_2$ &  7.3\\
    CO$_2$ &  4.26, 13.49*, 13.87*, 14.98, 16.18*, 16.78\\ 
     \hline\hline    
    \end{tabular}
    \begin{tabular}{p{\columnwidth}}
    *Emission features\\ 
    \citet{Gustafsson_1998,Matsuura_1999,Yamamura_2000,Sloan_2015}\\
    \end{tabular}
   \label{tab:molecules}
\end{table}

\section{Modeling} \label{sec:modeling}

 We use the one-dimensional Radiative Transfer (RT) modeling code DUSTY 2.07 \citep{Ivezic_Elitzur_1995,Nenkova_1999} to generate model SEDs for RT~Vir.  DUSTY solves the RT equation along the line of sight in one dimension based on a sequence of user-defined initial parameters \citep{Ivezic_Elitzur_1995}.
We compare the model DUSTY SEDs with observed spectral data for RT~Vir (Figure~\ref{fig:SED}) to generate as accurate a model of the dust environment as possible.

DUSTY allows us to adjust the nature of the central star, as well as the size, optical depth ($\tau_V$), inner dust temperature (T$_{\rm inner}$), density distribution and composition of the simulated dust shell. DUSTY provides us multiple ways to input the spectrum of a stellar photosphere. We can use a simple blackbody spectrum, or we can use a `naked' stellar atmosphere model. Model stellar atmospheres have the advantage that the strong molecular absorption bands that are used to classify cool stars are already present in the spectrum prior to passing light into our simulated dust shell.  As described in \S~\ref{sec:obs} we use a stellar spectrum template created from observations of M5III stars.

Our starting assumption for the radial density distribution is that it follows $r^{-2}$, which would reflect a constant mass-loss rate.
 Current dust formation models suggest that mass loss from AGB stars is influenced by several factors, particularly the pulsation cycle \citep[][and references therein]{Hofner_2016}, which may lead to periodic enhancements in dust density or clump formation \citep[e.g.,][]{Freytag_2023}. However, \citet{Villaver_2002B,Villaver_2002A} showed that hydrodynamic processes in the circumstellar shell erase such density structures, resulting in a $r^{-2}$ density distribution. Deviations from $r^{-2}$ will be described below. 

Our free parameters for modeling RT~Vir are, therefore, $\tau_V$ (related to the absolute dust density), T$_{\rm inner}$,  and the composition of the simulated dust shell. 
DUSTY includes both built in dust species, and an external library of complex refractive indices corresponding to a variety of materials, which we can supplement from the literature. 
While we used a range of silicate and oxide mineral optical constants, our best fitting models used the complex refractive indices of the following user-defined species: Updated ‘cosmic’ silicate and metallic iron from \citet{Speck_2015}, iron oxide from \citet{Henning_1995}, amorphous Al$_2$O$_3$  from \citet{Begemann_1997}, and crystalline Al$_2$O$_3$ from \citet{Pecharroman_1999}.

The crystalline Al$_2$O$_3$ refractive indices were generated from vibrational parameters of annealed hydrous aluminum oxides published by \cite{Pecharroman_1999}. The initial minerals were bayerite (monoclinic polymorph of gibbsite: $\beta$-Al(OH)$_3$) \& boehmite (AlO(OH)). \cite{Pecharroman_1999} heated each material at various temperatures and modeled their vibrational parameters, from which we were able to calculate the complex refractive indices of the resultant materials. The annealing experiments by \cite{Pecharroman_1999} generated final samples composed of multiple crystalline polymorphs of Al$_2$O$_3$ (e.g., $\alpha-, \delta-, \gamma-, \eta-, \theta$) in different combinations depending on the initial sample and the annealing temperature.

Our initial modeling assumed a single, continuous dust shell in which the composition is the same throughout, and using the M~Star template for the stellar photosphere. Using 
a mixture of cosmic glass, amorphous-Al$_2$O$_3$, Fe, and FeO, we could match the observational data shortwards of 
$\sim$22\,$\mu$m, but beyond that wavelength, 
the model begins to fail, being too low in flux and not matching the observed 28 and 32\,$\mu$m features. Increasing the amount of the coolest dust by modifying the radial density distribution led to unrealistic density profiles. 
\cite{Molster_2002} and \cite{Molster_2002b} showed that forsterite (Mg$_2$SiO$_4$, the Mg-rich end-member of the olivine series of minerals) can be associated with emission features at 28 \& 32\,$\mu$m. 
Therefore, to match the 28 and 32\,$\mu$m features 
we employed both forsterite, and fayalite (Fe$_2$SiO$_4$, the Fe-rich end member of the olivine series). Adding either of these minerals resulted in overestimation at shorter wavelengths and while features appeared near 28 and 32\,$\mu$m, the precise shapes of these features were too tall and narrow to match observations from ISO SWS. 

\cite{Sargent_2018} examined a set of AGB stars with similar 28--32\,$\mu$m features, and successfully modeled those features using the crystalline Al$_2$O$_3$ refractive indices from \cite{Pecharroman_1999}. They found that the 28--32\,$\mu$m features in their models were highly sensitive to the shape of the dust grains. 
By default, DUSTY assumes spherical dust grains. However, it is possible to simulate the effects of light interacting with dust of different shapes, by computing the absorption and scattering cross sections (C$_{\rm abs}$ and C$_{\rm sca}$) of  such dust outside of the DUSTY program. We have written a python program based on work in \cite{Min_2003, fabian_2001} \& \cite{Bohren_Huffman_1998} that generates the absorption and scattering cross sections for grains in a continuous distribution of ellipsoids (CDE;Dillon et al.\ 2025 {\it in prep.}). 
We tested each annealed sample from \cite{Pecharroman_1999} individually, and found that the products of bayerite annealed at 1273\,K  provide the best fit to the position and shape of features at 28 \& 32\,$\mu$m.  
Unfortunately, this material also resulted in flux overestimation at shorter wavelengths (like the olivine grains), which is consistent with expected behavior of crystalline Al$_2$O$_3$ \citep{Sloan_2003}. 

\begin{figure}
    \centering
    \includegraphics[width=\columnwidth]{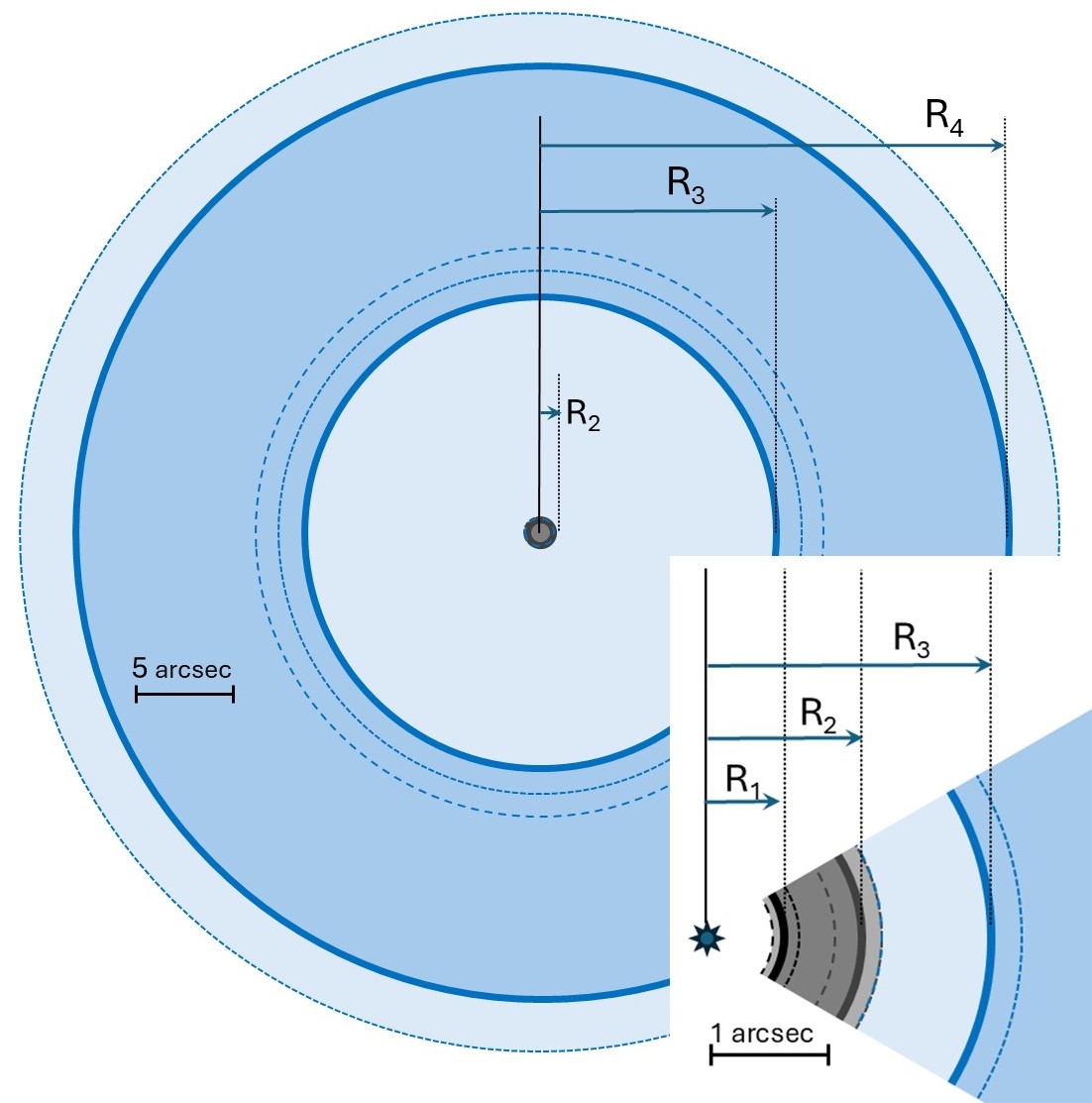}
    \caption{Schematic view of the two-shell DUSTY model. The inner dust shell is depicted in grey and black; the outer dust shell is depicted in blue. The thick arcs represent the nominal boundaries (inner and outer radius of each dust shell), while the dashed lines depict the minimum and maximum positions of those boundaries (longer dashes = minimum, shorter dashes = maximum). The darker grey and mid blue areas are the dust shells using the nominal boundary radii; the paler grey and blue mark the potential extent of each shell based on the minimum and maximum sizes. 
    R$_1$ is the inner dust radius for the inner dust shell;
    R$_2$ is the outer dust radius for the inner dust shell;
    R$_3$ is the inner dust radius for the outer dust shell;
    R$_4$ is the outer dust radius for the outer dust shell. Inset is an enlargement of the inner regions to show the inner dust shell details. Scale bars are based on the GAIA distance of 2.26\,pc.
    }
    \label{fig:schem}
\end{figure}

To model the 28 and 32\,$\mu$m features, we needed a gradient in composition with distance from the star. Introducing a gradient allowed us to include cool dust species that contribute emission features at 28 and 32\,$\mu$m without incurring too much emission at shorter wavelength.
Since DUSTY does not allow a user-input compositional gradient in a single model, we generated a two-shell model to simulate a change in composition over radial distance. 
We use the SED for the best-fitting single-shell model described above to replace the M~Star template as the input spectrum, and we ran another set of models. The output from this method returns a single model of two dust shells which can have different inner temperatures, different sizes, compositions, and optical depths. 
To avoid confusion, from this point forward we will use $\rm R_1, R_2, R_3, R_4$ to refer to the inner boundary of the inner shell, outer boundary of the inner shell, inner boundary of the outer shell and outer boundary of the outer shell respectively (see also Table~\ref{tab:inner_model_params} and Figure~\ref{fig:schem}).

\begin{table}
	\centering
	\caption{Parameters of the best fit DUSTY model. Sources for the optical properties of each dust species are listed below. Optical depth ($\tau_\lambda$) is evaluated at V-band (0.55\,$\mu$m).}
	\label{tab:inner_model_params}
	\begin{tabular}{lcc} 
	\hline
		Parameter & Shell 1 Value & Shell 2 Value\\
	\hline
            \multicolumn{3}{c}{Fractional Abundances}\\
        \hline
            Cosmic Sil$^{a}$ & 0.380 & 0.00\\
            Al$_2$O$_3~^{b}$  & 0.217 & 0.00\\
            ``Annealed Bayerite''$^{c}$ & 0.00 & 1.00\\
            FeO$^{d}$ & 0.163 & 0.00\\
            Fe$^{a}$ & 0.239 & 0.00\\
        \hline
            \multicolumn{3}{c}{Other Parameters}\\
        \hline
            T$_{\rm inner}$ (K) & 330${^{+30}_{-20}}$ & 94${^{+16}_{-5}}$\\
            ${\tau\,_V}$  & 0.16${\pm0.02}$ & 0.06${\pm0.02}$\\
            Inner Radius (AU)$^e$ & (R$_1$) 150${^{+20}_{-25}}$ & (R$_3$) 550${^{+65}_{-210}}$\\
            Outer Radius (AU)$^e$ & (R$_2$)300${^{+40}_{-50}}$ & (R$_4$) 10950${^{+1200}_{-4300}}$\\
		\hline
	\end{tabular}
 \\
    \begin{tabular}{p{\columnwidth}}
            $^a$~\protect\cite{Speck_2015}, $^b$~\protect\cite{Begemann_1997},
            $^c$~\protect\cite{Pecharroman_1999}; the annealed bayerite is actually a mixture of Al$_2$O$_3$ polymorphs, dominated by the $\alpha$- and $\theta$-structures, and does not contain any residual bayerite.
            $^d$~\protect\cite{Henning_1995}. $^e$~Assuming GAIA distance from Table~\ref{tab:RTVir}.\\
    \end{tabular}

\end{table}

DUSTY limits the input of absorption and scattering cross-sections to a single input file that comprises all the types of dust. Because of this limitation, the second shell models used CDE grains of a single size and composition for each model attempt. We first tried to fit features at 28 and 32\,$\mu$m with the crystalline silicates fayalite and forsterite; while we eliminated the overestimation of flux at short wavelengths, the 28 and 32\,$\mu$m features were still too sharp. Using the annealed bayerite from \cite{Pecharroman_1999} (with CDE grain) shapes provided the best fit to the 28 and 32\,$\mu$m in the ISO SWS spectrum. 

\begin{figure}
    \includegraphics[width=\columnwidth]{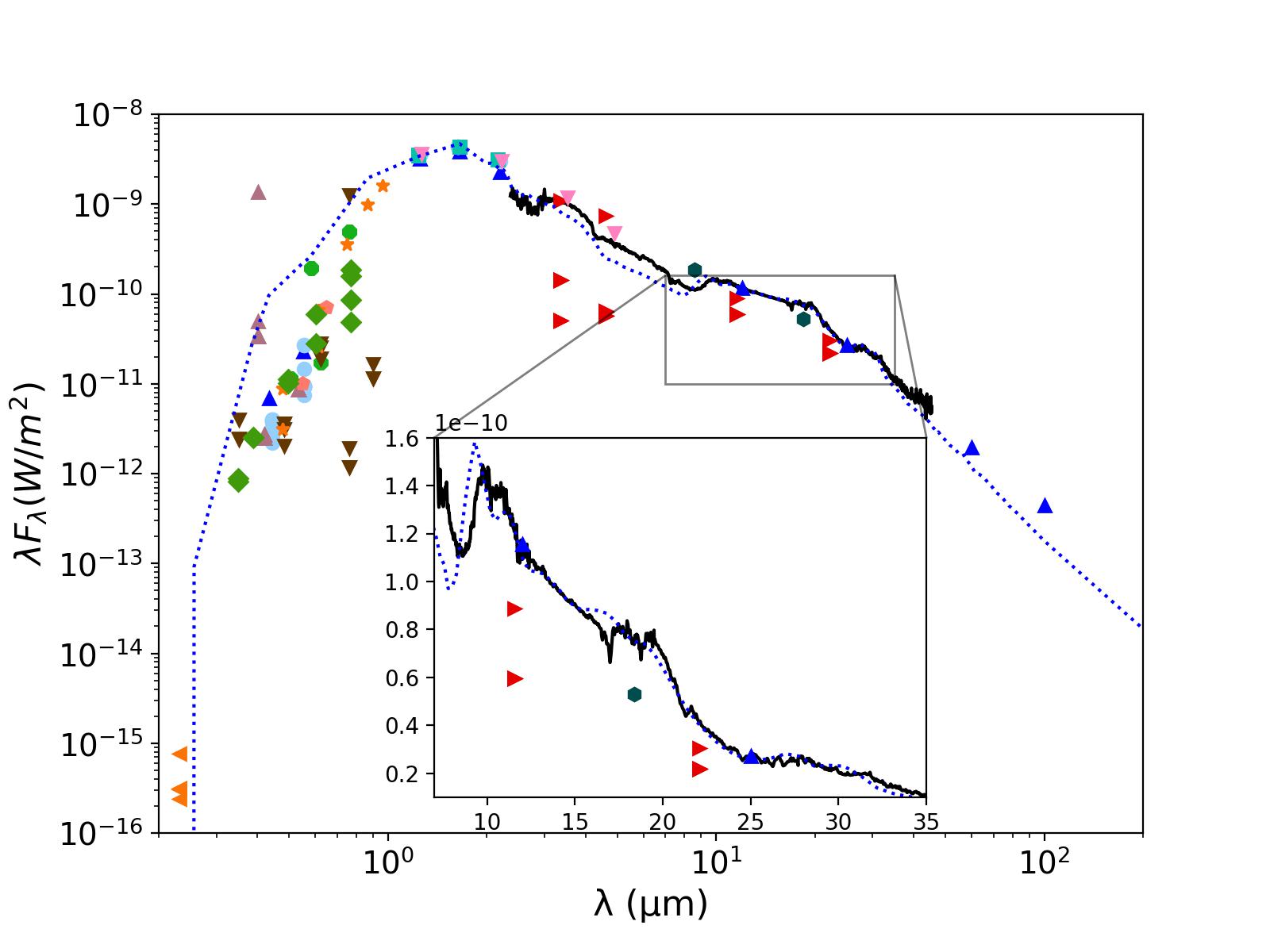}
    \caption{Best fitting DUSTY model SED for RT~Vir. Photometric points (colors) and ISO-SWS spectrum (black solid line) are identical to Figure~\ref{fig:SED}. Blue dotted line is the best fitting modeled SED with parameters as listed in Table~\ref{tab:inner_model_params}. Inset zooms into the mid-IR region (5--35\,$\mu$m) to show the details of the fit to the dust features}
    \label{fig:SEDModel}
\end{figure}

\begin{figure}
	\includegraphics[width=\columnwidth]{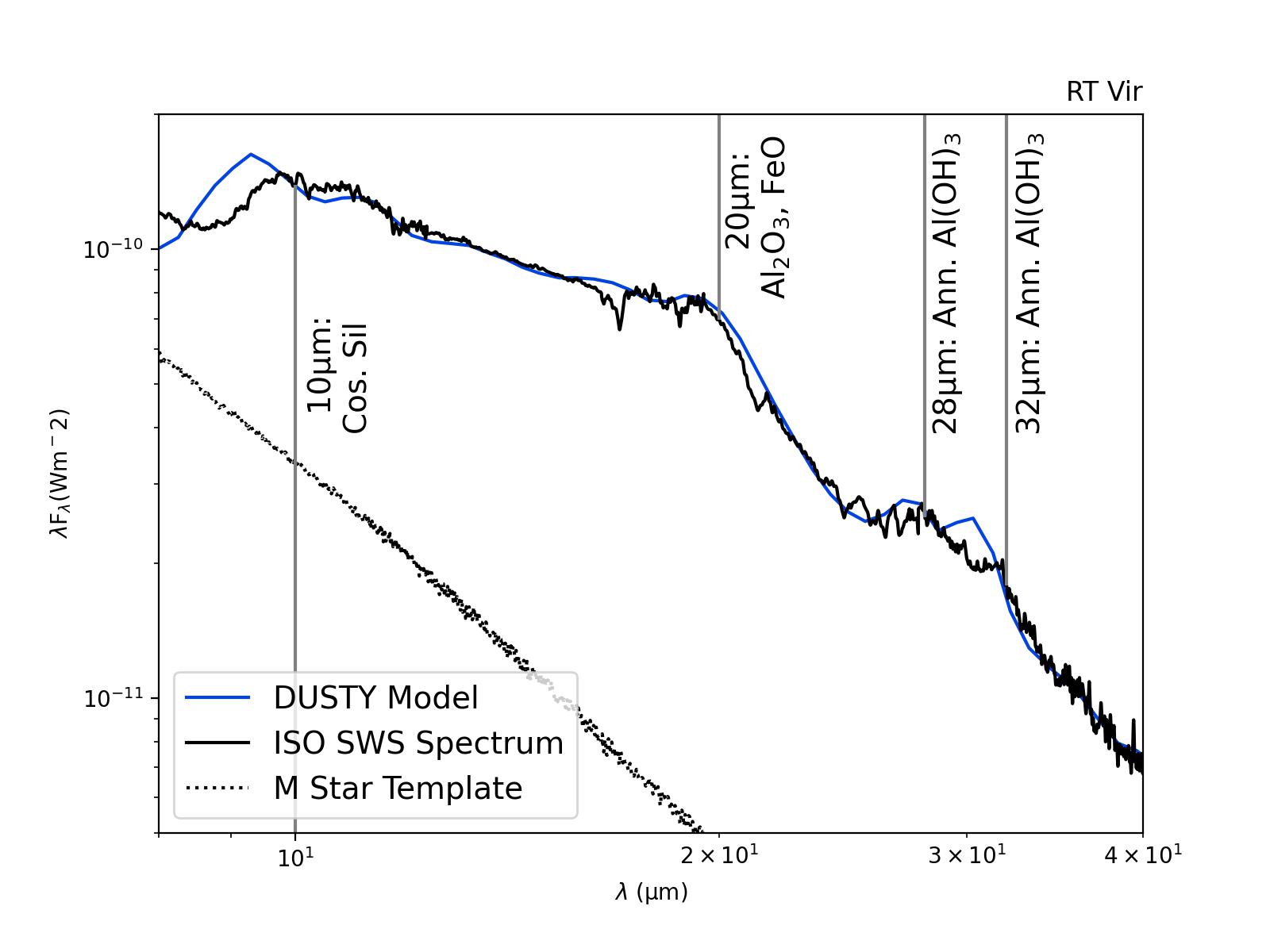}
    \caption{DUSTY model spectra for RT~Vir. Model spectra (blue) are fit to observed data (ISO SWS data in black) with important dust features highlighted by vertical grey lines (cosmic silicates, Al$_2$O$_3$, FeO, and annealed Bayerite). The black  dotted line shows the stellar template used to calculate the model.}
    \label{fig:ModelFigure1}
\end{figure}

Model fits were judged first by visually assessing the shape of the modeled SED  (see Figure~\ref{fig:SEDModel}~\&~\ref{fig:ModelFigure1}). Visual assessment is needed due to the difficulty of applying mathematical goodness of fit methods (e.g., $\chi^2$ fitting). Since DUSTY ingests only input parameters and does not physically simulate the system, fully automating DUSTY may return unphysical results, or potentially minimize $\chi$-squared by glossing over smaller features. Additionally, molecular features that are present in observation could skew the model output, and DUSTY does not model molecular gas features.  As discussed in \S~\ref{sec:obs}, the M~star template used to simulate the stellar photosphere is hotter than RT~Vir, resulting in a slightly brighter modeled SED at visual and UV wavelengths compared to observations. However, emission longward of 
$\mathrm{\sim 1\,\mu m}$ matches well. 

\begin{figure}
    \includegraphics[width=\columnwidth]{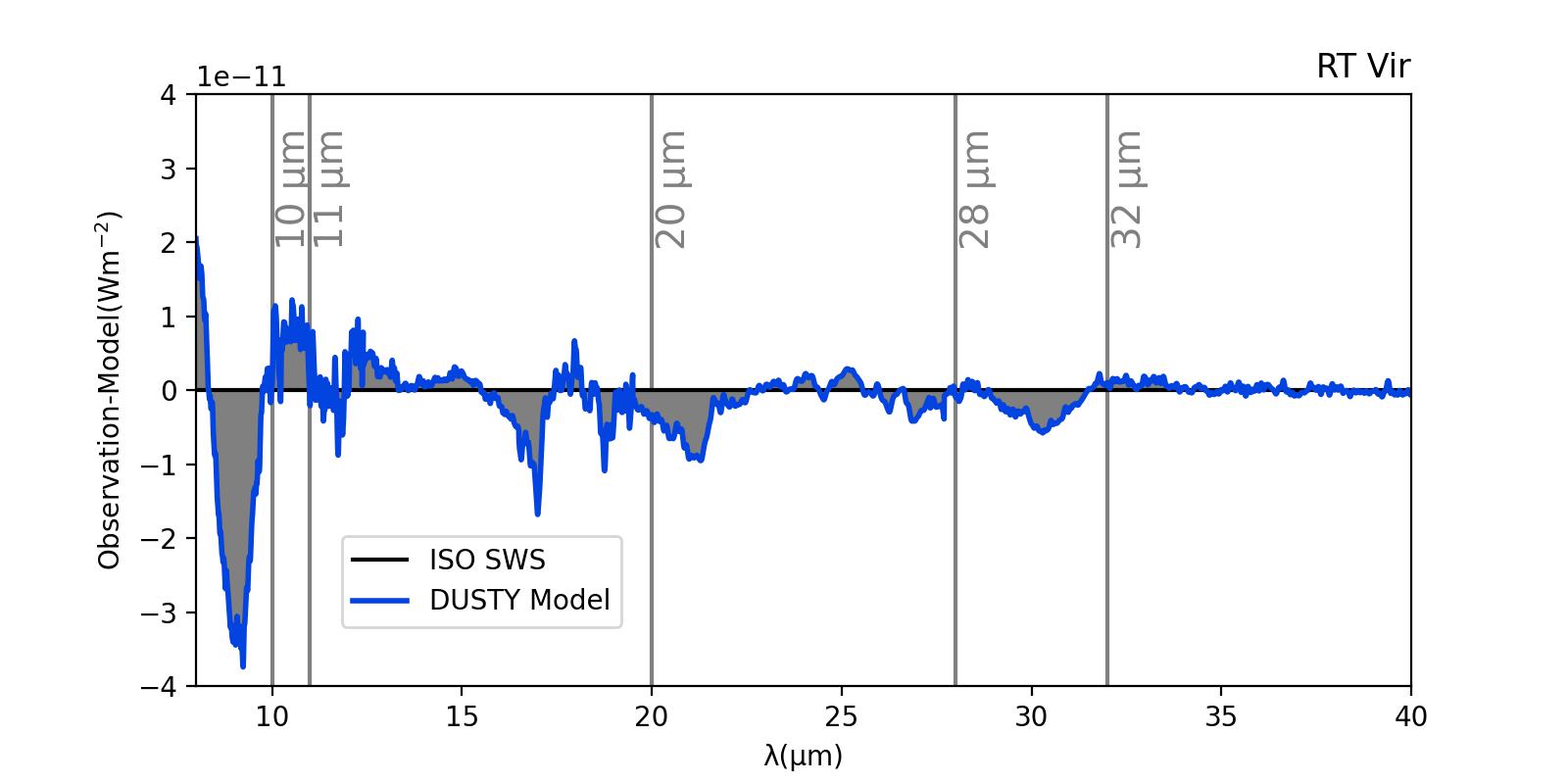}
    \caption{Residuals from the best fitting model above: Model(blue); ISO data (black). Vertical lines denote the same important features that are highlighted in Figure~\ref{fig:ModelFigure1}. The discrepancies are discussed in \S~\ref{sec:disc}}
    \label{fig:ModelRes1}
\end{figure}

To further assess the goodness of fit of our models, we subtract the model from the observed SED and plotted the residuals, shown in Figure~\ref{fig:ModelRes1}. This plot allows us to represent the ISO SWS spectrum as a straight line, and further assess where our model fits well, and where it does not. As mentioned above, DUSTY does not model the molecular features present in the SED. These features cause our model to over/underestimate emission, in the 5-9\,$\mu$m range, depending on the stellar template used. With this in mind, we will focus our modeling efforts on the part of the SED dominated by the infrared excess from dust at wavelengths greater than $\sim$8\,$\mu$m.

\section{Model Results} \label{sec:results}

Our best fitting DUSTY model for RT~Vir is shown in 
Figures~\ref{fig:SEDModel} \&~\ref{fig:ModelFigure1} with residuals in Figure~\ref{fig:ModelRes1} and the DUSTY model parameters are given in 
Table~\ref{tab:inner_model_params}. These parameters indicate the presence of
a highly extended, optically thin dust shell consisting of two distinct layers as shown schematically in Figure~\ref{fig:schem}. 
Figure \ref{fig:ModelFigure1} shows the model fit to the ISO SWS spectrum focused on the star's IR dust excess ($8\mu m\leq\lambda\leq40\mu m$). Figure \ref{fig:ModelRes1} shows the residual plots for the same model shown in Figure \ref{fig:ModelFigure1} over the same range of wavelengths. This figure was used to assess the goodness of fit for our model SED, and highlights where the model either over or underestimates emission. The grey lines in both Figure~\ref{fig:ModelRes1} \&~\ref{fig:ModelFigure1} indicate the positions of well known dust features. 
In our model of the inner shell, the largest fraction of dust is made up of cosmic silicates with smaller fractions of Al$_2$O$_3$, Fe, and FeO. The dust temperature, $T_1$  at the inner radius, $R_1$, of the dust shell is 330\,K. In the model of the outer shell, composition is held constant at 100\% annealed bayerite \citep[ from][]{Pecharroman_1999}, which consists of a mixture of Al$_2$O$_3$ polymorphs, with $\theta$- and $\alpha$-Al$_2$O$_3$ having the greatest fractional abundance in the sample. Inner temperature, $T_3$  at the inner radius, $R_3$, of this shell is 94\,K.  This dust is particularly cool; and since DUSTY calculates dust shell radius using the inner dust temperature (T$_{inner}$) the calculated shell sizes are large as a result  (see \S~\ref{sec:pt} for complete discussion of pressure-temperature parameter space for this star).

It is worth noting that while the nominal positions of the dust layers leaves a gap of a few hundred AU between the inner and outer shells, if we assume the maximum extent for the inner shell (R$_2$ is at maximum) and the minimum distance for the inner boundary of the outer shell (R$_3$ is at minimum), then R$_2$ = R$_3$ and we have a continuous dust shell, although the composition changes at this point (see Figure \ref{fig:schem}).

\section{Discussion} \label{sec:disc}

\subsection{Dust Composition}

Based on the results of our RT modeling, RT~Vir has circumstellar dust that is distant and cool. The dust cloud itself is relatively optically thin, with $\tau_{V} \approx 0.16$.

We successfully model the overall shape of both the SED and most prominent dust features at 10, 11, 18-20, and 28-32\,$\mu$m, as well as the broad emission 'bridge' between $\sim$10 and 20\,$\mu$m. In O-rich systems, emission at 10 and 18\,$\mu$m is associated primarily with warm silicate grains. Other common features at 11\,$\mu$m, 13\,$\mu$m, 20\,$\mu$m are associated with  two polymorphs of Al$_{2}$O$_{3}$ and FeO \citep{Speck_2000, Sloan_2003}. Features at 28-32\,$\mu$m are often associated with crystalline grains, and in this case best modeled by a mixture of crystalline Al$_2$O$_3$ polymorphs. 

Studies of dust around O-rich AGB stars have revealed multiple classes of AGB stars \citep[e. g.][]{Little-Marenin_Little_1988, Sloan_Price_1998, Speck_2000}. RT~Vir, however, does not fit cleanly into any of these classifications. The dust features for RT~Vir are broad and don't present clear peaks, but remain different from the classic ``broad'' feature described by \citet{Little-Marenin_Little_1990,Speck_2000}. In spite of this notable difference, we have produced a successful model using the commonly expected astrominerals described above. The key differences for RT~Vir's model parameters are the cool dust and vast shell size.

\subsection{Large extent of the modeled dust shell} \label{sec:extent}

\begin{figure}
    \includegraphics[width=\columnwidth]{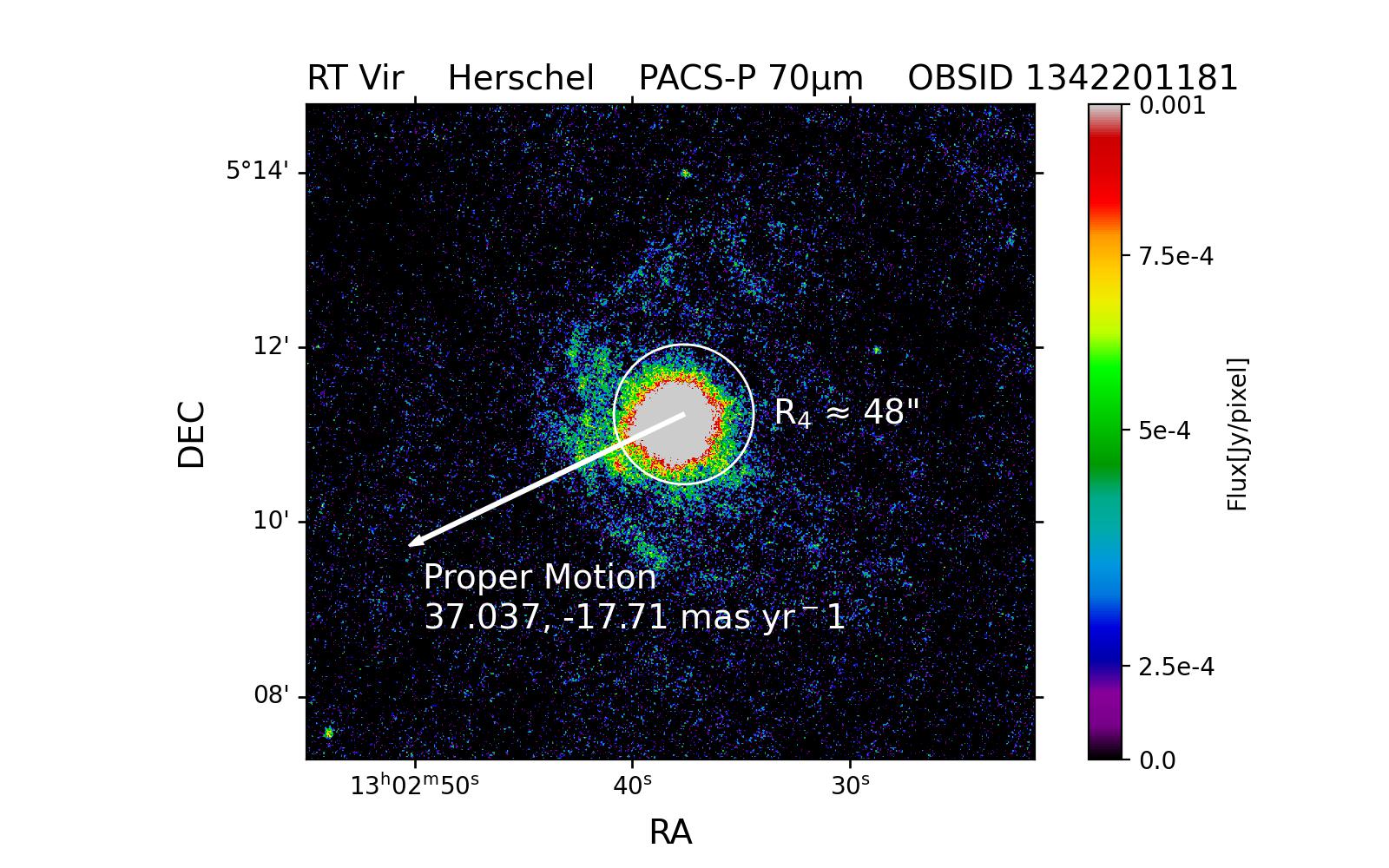}
    \caption{Herschel PACS 70\,$\mu$m image of RT~Vir, showing the fermata (\fermata) shaped bow-shock emission that extends to $\sim$3$'$ away from the central star. The white arrow indicates the direction of the proper motion of RT~Vir. The white circle marks the outer extent of our modeled dust shell (R$_4 \approx$ 48")}
    \label{fig:herschel_postcard}
\end{figure}

The large extent of the model may seem surprising but is consistent with imaging observations of RT~Vir at multiple wavelengths.
The Herschel Space Observatory PACS instrument  observed RT~Vir in the 70 and 160\,$\mu$m bands in 2010 as part of the MESS (Mass-loss of Evolved StarS) Herschel key program \citep{Groenewegen_2011}. The 70$\mu$m-image is shown in Figure~\ref{fig:herschel_postcard} depicting far-IR emission up to $\sim 3'$ away. \cite{Cox_2012} modeled AGB stars from the MESS Key Program and found that RT~Vir has a wind-ISM interaction region with pronounced Kelvin-Helmholtz instabilities, and a general ``fermata'' shape or \fermata~ (this is a musical symbol that strongly resembles the bow shock and star combination). \citet{Cox_2012} modeled the standoff-distance between this bow shock and the star at 2.6\,$'$. The dense bow-shock emission at 70\,$\mu$m shows an angular separation of 2.9\,$'$ at the extreme edge of the densest emission along the bow shock (see Figure~\ref{fig:herschel_postcard}) . 
This fermata shape is indicative of having a leading bow-shock formed as the stellar wind collides with the ISM in the direction of the proper-motion of the star \citep[see, e.g.,][]{Wareing_2006}. 

The extent of the RT~Vir dust shell in these observations translates to a shell size of $\sim$40,000\,AU assuming the GAIA distance from Table~\ref{tab:RTVir}. 
RT~Vir has also been found to be extended at other wavelengths.
The 2~Micron All Sky Survey \citep[2MASS;][]{Skrutskie_2006A} took images in the J, H, and K bands; the K-band image shows extended gas and dust emission $\sim 2'$ ($\sim$27,000\,AU) away. 
\cite{Sahai_2023} used GALEX observations of RT~Vir to show the bow shock and astrotail of the star align with its proper motion vector. Thus RT Vir is extended in UV wavelengths.
RT~Vir has also shown extended emission in images from AKARI \citep{Ueta_2018}, and the Spitzer Space Telescope (unpublished). All these observations suggest dust and gas extends at least twice as far as our modeled outer dust shell. With the most recent distance estimates to RT~Vir \citep[$\sim$226$\pm$7\,pc][]{Zhang_2017, Andriantsaralaza_2022}, the outer extent of our modeled dust shell should appear around 48$''$ ($\sim0.8'$). Our models are therefore not unreasonable.

\subsection{Drift Velocity}
\label{sec:drift}

For what follows in \S~\ref{sec:pt}--\S~\ref{sec:epochs}, we need to discuss the outflow speeds, which impact calculations of timescales and pressure-temperature regimes.
Although we use the gas outflow velocity as measured from CO observations \citep{Loup_1993,Olofsson_2002}, this may not be identical to the dust outflow velocity. 
The prevailing hypothesis for outflows from AGB stars is that dust forms in the stellar atmosphere and is propelled outward by radiation pressure from the star’s luminosity. As the dust accelerates, it transfers momentum to the molecular and atomic gas through collisions. Consequently, the gas outflow velocity at large distances from the star depends on the stellar luminosity ($L_\star$), the dust-to-gas ratio, the mass-loss rate ($\dot{M}$), and the properties of the dust grains.
The difference between the gas velocity (\vg) and the dust velocity (\vd) is known as the drift velocity (\vdr).

Direct measurements of dust velocity are challenging because the spectral features are too broad to detect Doppler shifts. In contrast, molecular gas, typically observed through CO in radio wavelengths, allows for the determination of both Doppler broadening and the gas mass-loss rate.
In general, it is assumed that the drift velocity is not huge and that the dust outflow velocity is similar to gas outflow velocity \citep[e.g.,][]{Hofner_2018}, the reality depends on the optical depth of the dust shell. For optically thick dust shells, the drift velocity is expected to be negligible \citep{Gail_Sedlmayr_1985}. However, the case for optically thin dust shells, like that of RT Vir, is less clear. Direct measurements of dust velocity were achieved by \citet{Dougados_1992} using dust-scattered photospheric emission lines in a single object (Frosty Leo, post-AGB object) and found that the dust velocity varies but is consistently 10--20\kms slower than the CO molecular gas velocity, however, they did suggest this could be an artifact of the radiative transfer process. Meanwhile, \citet{Goldreich_1976} modeled dust around the more optically thick OH-IR star type and found the drift velocity to be 9\kms (dust faster than gas). And \citet{Berruyer_1983} developed a two-fluid model that produced dust velocities of $\sim$2--4\,km\,s$^{\rm -1}$ and gas velocities of 1-2km/s. More recent models have been developed that suggest much higher drift velocities. \citet{Sandin_2023} included dust-gas drift into their models for mass loss from O-rich AGB stars and found \vdr~could be as high as 300\kms, while \citet{Zargarnezhad_2023} used a dust-gas-drift modeling technique to investigate the formation of clumps and inhomogeneities in AGB star shells and found \vdr $\approx$30\kms (for \vg $\approx$11\kms). It should be noted that \citet{Kwok_1975} stated that \vdr $\gtrsim$ 20\kms would lead to dust grain destruction via sputtering.

\citet{Habing_1994} studied whether the gas outflow velocity could be used to estimate the gas-to-dust ratio, and came up with an equation for the drift velocity in terms of stellar luminosity, mass loss rate, gas velocity and the average radiation pressure efficiency ($\left<Q_{\rm rp}\right>$, which, in turn, is determined by the nature of the dust grains in terms of their size and opacity). This is identical to the equation derived by \citet{Goldreich_1976}:

\begin{equation}
   v_{\rm drift}^2 = \frac{\left<Q_{\rm rp}\right> L_\star}{\dot{M}c}v_{\rm gas}
   \label{eq:qrp}
\end{equation}

Since the mass-loss rate and luminosity are both expected to increase during the AGB phase \citep[see e.g.][]{vanLoon_2005,Vassiliadis_1993}, these should not have a huge effect on the drift velocity, but the nature of the dust grains may change and thus the drift velocity may vary with time.
Calculating $\left<Q_{\rm rp}\right>$ is not straight forward but we can approximate using functions in \citet{Draine_1981} which assume either carbon or “astronomical silicate” grains. For silicates we get a $\left<Q_{\rm rp}\right> \approx 0.1$ for 1\,$\mu$m size grains which is consistent with the values calculated from forsterite \citep{Gilman_1969} and adopted by \citet{Berruyer_1983}. Smaller grain sizes give rise to smaller drift velocities.

Based on the opacities listed in Table~1 of \citet{Bladh_2012}, the wavelength-averaged opacity of Al$_2$O$_3$ is very similar to that of non-iron-bearing silicates. In comparison, metallic iron and iron-bearing olivine are approximately 30 and 60 times more opaque, respectively, with amorphous carbon being even more opaque.
 Since $\left<Q_{\rm rp}\right>\approx 0.1$ is valid for all non-iron bearing minerals, and most of our dust is not iron-bearing or carbonaceous, we can apply Equation~\ref{eq:qrp} and get a dust drift velocity of ~7km/s. 
  Our model also includes metallic iron grains, which would exhibit higher drift velocities; however, the overall dust drift velocity is reduced due to the predominance of much smaller grains. We assume an MRN grain size distribution \citep[see][]{Mathis_1977}, which has a maximum grain size of 0.25\,$\mu$m and follows a power law, resulting in most grains being significantly smaller (5\,nm and up). In addition, since we are examining dust in its earliest stages of formation, we expect the predominance of many small grains. For a discussion comparing grain sizes relevant to AGB stars, see \citet{Speck_2009}.

Given the variations in modeling and observational results, and that our model assumes an MRN grain size distribution we assume that $v_{\rm dust} = v_{\rm gas}$, but with $\pm$5\kms to account for potential errors in our timescale calculations.

\subsection{Mass-Loss Timescales}
\label{sec:timescale}

While the timescale associated with cooler inner dust temperatures ($\sim$100s of years) have periods too long to measure on human timescales, we can observe the shells of dust around the star, and examine the history of its outflow. 

Using the GAIA distance of 226\,pc \citep[][and Table~\ref{tab:RTVir}]{Zhang_2017} and the CO gas expansion velocity ($v_{gas}$) of 11.3\,\kms from \cite{Loup_1993}, we have calculated the timescales associated with the two distinct dust shells, which are listed in Table~\ref{tab:cartoon_timescales}. The errors quoted are based on those from Table~\ref{tab:inner_model_params} plus allowing for a drift velocity $v_{\rm drift}$ of 5\kms, which we discuss in more detail in \S~\ref{sec:drift}. 

\noindent
\begin{table}
\centering
\caption{Timescales for the age of each shell boundary based on CO wind velocities calculated in \protect\cite{Loup_1993}}    
\label{tab:cartoon_timescales}
\begin{tabular}{lcccc}
\hline
Boundary & Age$^a$ & Temp$^b$ & Radial       & Angular\\
               & (yrs)   & (K)      & Distance$^b$ & size$^c$\\
               &         &          & (AU)         &  \\
\hline
\multicolumn{5}{c}{inner shell}\\
\hline
$\rm R_1$ &  63$^{+72}_{-25}$  & 330 & 150 & 0.66$''$\\    
$\rm R_2$ & 125$^{+129}_{-53}$ & 238 & 300 & 1.32$''$\\
\hline
\multicolumn{5}{c}{outer shell}\\
\hline
$\rm R_3$ & 229$^{+231}_{-131}$    & 94 & 550 & 2.42$''$\\
$\rm R_4$ & 4590$^{+4544}_{-2658}$ & 34 & 10950 & 48.4$''$\\
\hline
\end{tabular}
\begin{tabular}{p{8cm}}
$^a$ based on $v_{gas}$ from Table~\ref{tab:RTVir}; $^b$ From Table~\ref{tab:inner_model_params}; $^c$ based on the GAIA distance of 226\,pc \citep[][and Table~\ref{tab:RTVir}]{Zhang_2017}.\\
\end{tabular}
\end{table}

Table~\ref{tab:cartoon_timescales} shows that the size of our modeled dust shells correspond to several hundred years of different dust outflows. These separate epochs of dust formation could represent two periods of dusty outflow followed by an extended period of no or low dust production leading up to the present. It should also be noted that the dust accelerates as it moves away from the star, reaching a terminal velocity some time later \citep[e.g.][]{Habing_1994}. This means that the ages calculated here, which assume a constant outflow velocity, are minimum age estimates for each dust shell.

The timescale for the formation of discrete dust shells suggested by our models is  of the order of hundreds of years. This matches neither the pulsation timescale (hundreds of days) or the thermally pulsing timescale ($10^4-10^5$ years). However there are several {\em carbon-rich} AGB and post-AGB systems that exhibit discrete dust shells on similar times scales to those from our models.
LL~Peg (an ``extreme’’ C-rich  AGB star, also known as AFGL~3068) has a stunning pinwheel structure of dust density that has a timescale of $\sim$800\,yrs between dust shells \citep{Morris_2006, Mauron_Huggins_2006}. Similarly,
CIT~6 has a  pinwheel structure observed in the molecular gas (HC$_3$N), which has timescales for mass-loss changes in the 400--1200\,yr range \citep{Claussen_2011}. A similar multi-ringed shell was observed in molecular gas around C-star, V~Hya, with timescales in the hundreds of years \citet{Sahai_2022}. Meanwhile, the archetypal carbon star, IRC+10216 was observed to have discrete dust rings in V band too \citep{Mauron_Huggins_2010}. Post-AGB object AFGL~2688 \citep[a.k.a.\ the Egg Nebula;][]{Sahai_1998} also exhibits episodic mass loss, with dust shells that formed every 150-450\,yrs, for a duration of 75-200\,yrs.
Finally, \citet{Randall_2020} showed that O-rich AGB star, GX~Mon, has a broken spiral shell structure extending out to 4000\,AU, as measured in molecular gas.
Although there is a paucity of similar structures around O-rich mass-losing stars, the C-rich object demonstrate that AGB stars often have dust shells with timescales in the hundreds of years. And GX~Mon demonstrates that similar structures are possible for O-rich AGB stars. 

\subsection{Pressure-Temperature Space}
\label{sec:pt}

\begin{figure}
    \centering
    \includegraphics[width = 4in]{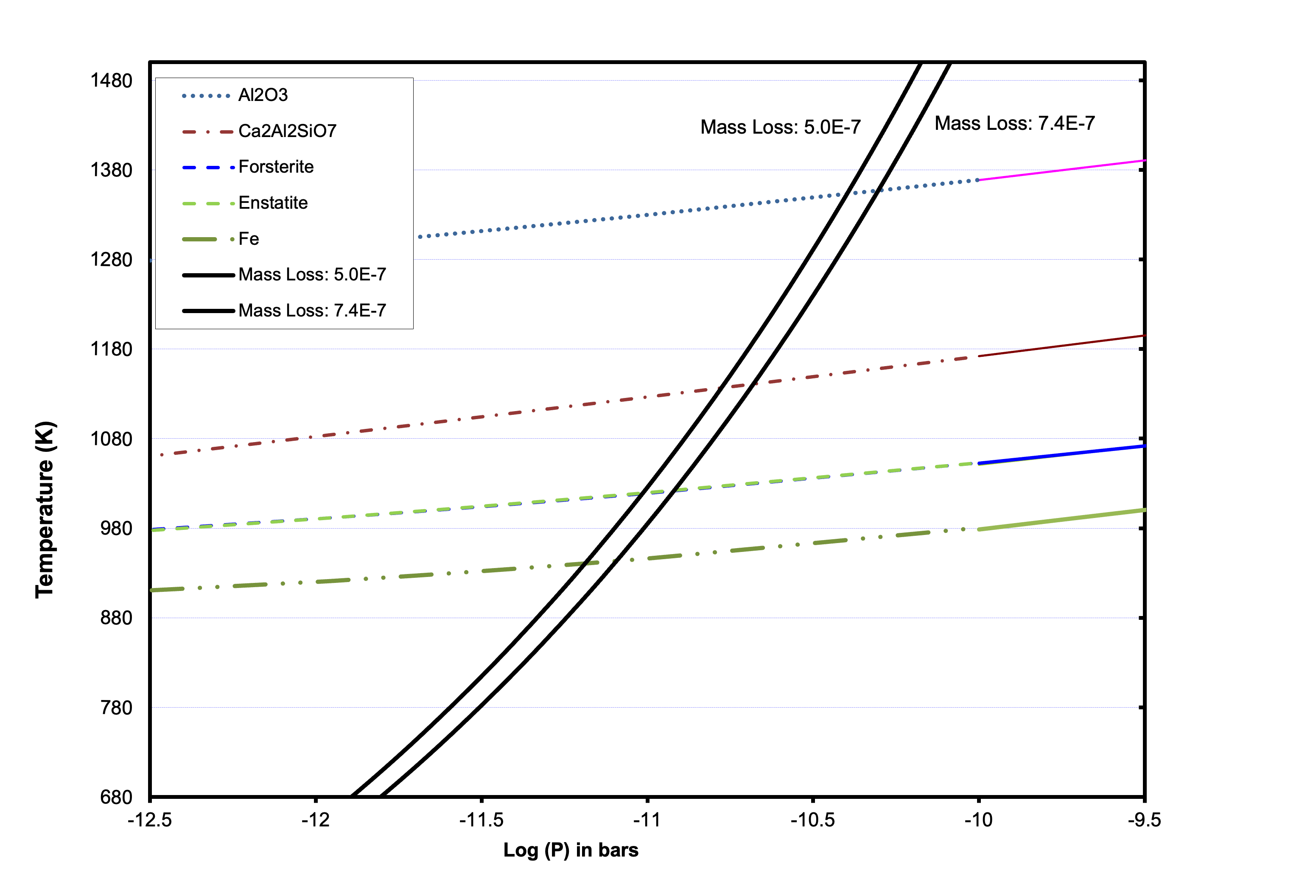}
    \caption{Pressure-temperature diagram for the space around RT~Vir. The dark lines with labels show the PT-space occupied by RT~Vir based on different calculations of the expansion velocity and dust mass-loss rate \protect\citep[M$_\odot$yr$^{-1}$][]{Olofsson_2002, Loup_1993}. Dashed and dotted lines indicate the PT conditions required for certain minerals to condense out of the circumstellar material, extrapolated from data published in \protect\cite{Lodders_Fegley_1999}.}
    \label{fig:PT_space}
\end{figure}

Figure~\ref{fig:PT_space} shows the Pressure-Temperature (PT) diagram for RT~Vir, where the pressures and temperatures are calculated using mass-loss rate ($\dot{M}$) and CO gas expansion velocities ($v_g$) listed in Table~\ref{tab:RTVir}. The calculations follow the methodology from \cite{Speck_2009}. 
Where the dark RT~Vir lines intersect dashed and dotted stability lines for condensates indicates the temperature at which said condensates should first become stable (assuming thermodynamic equilibrium). We can see that different minerals should become stable at different pressures and temperatures within the PT space bounded by the mass-loss rate range. The lowest temperature silicate should become stable above $\sim$1000\,K and metallic iron above $\sim$900\,K. However, our models suggest far cooler dust temperatures, near $\sim$330\,K. 

In \S~\ref{sec:formation}, we will discuss dust formation in more detail but it is important to note that the dust formation temperatures predicted above, which  assume thermodynamic equilibrium, should be considered an upper limit.
Dust condensation is expected to occur in the 800--1000\,K range, and this can be accommodated in models for most AGB stars. However, our models suggest that dust around RT~Vir is distant, and cool ($\sim$ 330K at the innermost boundary). Since our models indicate the presence of cool, distant dust shells rather than compact, warm dust, this may suggest a slowdown in dust production at RT~Vir.

\cite{Marengo_2001} found that many AGB stars, particularly non-Mira variables, are best characterized by cooler inner dust temperatures. As an SRb variable, RT~Vir may follow this trend.

If RT~Vir was currently losing mass, one might expect to see hotter dust. Using the stellar luminosity and the theoretical condensation temperatures for common dust species (see Figure~\ref{fig:PT_space}) we can calculate the distance at which dust should form as follows:

\begin{equation}
    \label{eqn:PTDist}
    r = \sqrt{(1 - A) * \frac{L_*}{16 \pi \sigma T_{cond}^4}}
\end{equation}

Where $A$ is the albedo of the dust, $L_*$ is the stellar luminosity in Watts, and $\mathrm{T_{cond}}$ is the condensation temperature of the dust. If we take $L_* = 5000\mathrm{L_{\odot}}$ \citep{Brand_2020}, $T_{\rm inner} = 1100\,K$, and $A = 0.5$ we can get a rough estimate of the dust formation distance: $\mathrm{r = 4.7\times10^7km = 3.14AU}$. 
Evidently, this is far closer to the stellar surface than our modeled shells $\mathrm{(r \approx 2.23\times10^{10}km = 149\,AU)}$. 
The temperature $T_{\rm inner} = 1100\,K$ was selected based on Figure~\ref{fig:PT_space}, which indicates that Al$_2$O$_3$ is expected to condense at $\sim 1350$\,K for RT Vir assuming thermodynamic equilibrium. However, as will be discussed in \S~\ref{sec:formation},  dust nucleation probably requires the outflowing gas to achieve supersaturation before dust can nucleate. To account for this effect, a slightly lower temperature — several hundred degrees cooler — was selected. 
The range of potential condensation temperatures $T_{cond}$ depends on the composition of the gas, the mass-loss rate, the expansion velocity and the temperature of the star. For the common Mg-rich/iron-poor silicates (forsterite, enstatite), the thermodynamic equilibrium temperature does not drop much below 1000\,K. Even when considering supersaturation requirements,  $T_{cond}$ or $T_{\rm inner}$ should be above 800\,K. Our modeled $T_{\rm inner}$ is only 330\,K

\subsection{Dust Formation}
\label{sec:formation}

There are multiple competing (or at least different) theoretical models of dust formation in the stellar outflows around AGB stars. One, and potentially the most intuitive is a model of thermodynamic equilibrium condensation. \cite{Lodders_Fegley_1999} have calculated the condensation temperatures and pressures of various silicates as well as Al-bearing minerals thought to be present in C \& O-rich AGB outflows. Figure~\ref{fig:PT_space} uses these condensation properties for Al$_2$O$_3$, calcium aluminosilicate, enstatite, forsterite, and iron. It should follow then, that when the pressure-temperature space around an AGB star intersects the condensation pressure and temperature for a mineral, dust grains of that substance should start condensing so long as the elements that compose it are suitably abundant.

In contrast to forming at thermodynamic equilibrium dust grains may also form under conditions of supersaturation \citep[e.g.,][]{Nuth_Hecht_1990, Gail_Sedlmayr_1999}. In this case the outflowing gas cools, reaching a state of supersaturation before slow-forming mineral condensates can produce significant dust emission \citep{Gail_Sedlmayr_1999}. Al$_2$O$_3$, having the highest condensation temperature in O-rich AGB outflows, will have the longest time to form grains that can become the nuclei for silicate mantles to grow atop \citep{Gail_Sedlmayr_1998}.

 \cite{Nuth_Hecht_1990} suggested that dust grains form around AGB stars by quickly condensing out of the supersaturated circumstellar vapor into 'chaotic silicates'. Within this chaotic grain, Si, Al, and other metals would not be fully oxidized, and since Al has the highest reduction potential, it would oxidize preferentially, even converting $\mathrm{Al + SiO \rightarrow AlO + Si}$ \citep{Stencel_1990}. It follows then, that in an O poor environment, Al will oxidize first, and may remove most of the free O in the environment. In this case, there will be less SiO to form silicate grains or mantles atop Al$_2$O$_3$, resulting in a lack of the associated emission at 10 and 20\,$\mu$m.

\subsection{Epochs of Dust Formation}
\label{sec:epochs}

With these hypotheses of dust formation in mind, we can turn our attention back to our modeled dust shells around RT~Vir. Our two modeled dust shells are likely indicative of two separate epochs of dust formation in the outflow of RT~Vir. Since the outer shell is modeled using crystalline material, it likely formed closer to the star, and cooled slower, giving the Al$_2$O$_3$ time to anneal in the outflow. This could occur if the grains formed at thermodynamic equilibrium, or in a supersaturated gas, so long as the grains stayed at a high enough temperature (1100--1300\,K, which include the temperatures at which \cite{Pecharroman_1999} annealed their bayerite samples) for long enough to anneal \citep[e. g.][]{Gail_Sedlmayr_1999}. The inner shell likely cooled faster, and could form amorphous grains, or amorphous mantles. In this case it is less likely that all grains formed at thermodynamic equilibrium, and more likely that grains condensed directly out of a supersaturated outflow at lower temperatures, or that amorphous mantles grew on top of crystalline or amorphous nuclei \citep{Stencel_1990, Gail_Sedlmayr_1999}. In this case, the composition and crystallinity of the dust shells depend explicitly on the mass-loss rate of RT~Vir, which governs the PT environment in the outflow (see Figure~\ref{fig:PT_space}). 

Alternatively, if the dust around RT Vir formed primarily as chaotic grains \citep[e. g.][]{Stencel_1990}, then the C/O ratio would drive the dust distribution and makeup over the stellar mass-loss rate. The enhanced abundance of Al$_2$O$_3$ compared to cosmic silicate (especially in the outer shell) is best explained as a function of high C/O ratio. If Al$_2$O$_3$ dust condenses before other dust species then the abundance of free oxygen to form silicates would be depleted. Additionally, in chaotic grains Al$_2$O$_3$ may form preferentially due to its high reduction potential compared to that of SiO$_2$ \citep[e.g.][]{Stencel_1990}. In either case, since C/O is high, Al$_2$O$_3$ may reduce the amount of free O atoms in the outflow, enough to prevent other O-bearing minerals (FeO, SiO$_2$, etc...) from condensing in great abundance. In the case of chaotic grain formation, a decrease in the C/O ratio could explain the shift in grain chemistry between the shells of our models.

One phenomenon that can decrease an AGB stars C/O is carbon burning in the star's envelope. In more massive AGB stars (M $>$ 4.0\,M$_\odot$), the bottom of the outer stellar envelope can begin the CNO cycle, consuming carbon during the third dredge up period in a process called hot bottom burning \citep[HBB; described in ][]{Boothroyd_1993}. As carbon is consumed in the stellar envelope, the C/O ratio may decrease over time, which is not what we expect for lower mass AGB stars. This decrease in C/O may account for the changes between dust shells in our models of RT~Vir. If the outer shell formed when the circumstellar material had C/O greater than when the inner shell formed, it may suggest that the outer shell formed as chaotic silicates. 
Then, as C/O decreased as HBB occurred in the star, increased O abundance would encourage the formation of silicates and other oxides.

However, this can only be true if RT~Vir is indeed greater than $\sim$4M$_\odot$. We  found one estimate of RT~Vir's mass from \cite{Yates_2002} at 1.5M$_\odot$ (too small to initiate HBB). However, no description of a method or citation for this number are given, and thus calls into question the reliability of this mass estimate. If RT~Vir is less than $\sim$4M$_\odot$, considering the C/O ratio as the determining factor in dust species formation would be difficult to justify. In this case the mass-loss rate would drive the change in composition between the outer and inner shells. In this case, Al$_2$O$_3$ in the outer shell may have formed during a period of low mass-loss rate, when the circumstellar envelope had enough pressure to form dust, but was too hot for anything other than Al$_2$O$_3$ to condense. After the circumstellar material cooled enough for other materials to condense, the pressure may have been too low to support significant formation of silicate mantles on Al$_2$O$_3$ grains or any other oxides. If the mass-loss rate increased between the formation of the outer and inner shells, the pressure could have stayed high enough to form silicate mantles and other oxides at temperatures below $\sim$1100K.

\newpage

\section{Conclusions}
\label{sec:conc}

We have presented DUSTY modeling of the circumstellar dust of O-rich AGB star RT~Vir. Our modeling was matched visually to IR SEDs for RT~Vir. Our models suggest a pair of distant \& cool shells of dust with low optical depths. The inner, younger shell in our best fit model has composition composed primarily of silicates and Al$_2$O$_3$ with additional components of FeO and metallic Fe. The outer, older shell is modeled entirely using a mixture of Al$_2$O$_3$ polymorphs, primarily 
$\theta$-Al$_2$O$_3$. From this modeling we can assert the following:
\begin{enumerate}
    \item RT~Vir's circumstellar material is cold and distant compared to other O-rich AGB stars.
    \item The distances from the star to our modeled dust shells are consistent with imaging at 70\,$\mu$m via Herschel.
    \item There are two scenarios that could explain our results at RT Vir:
    \begin{itemize}
        \item The differences in composition between the    outer (older) and inner (younger) shells of dust suggest a change in the mass-loss rate. This would be consistent with theories of TE or supersaturation dust formation, which are driven by PT conditions over C/O ratio \citep[e.g.,][]{Lodders_Fegley_1999, Gail_Sedlmayr_1999}.
        \item The compositional differences between the outer and inner shells of dust could be due to a decrease in C/O between the formation of the shells, which could be consistent with the formation of chaotic silicates  which is driven by C/O more than PT conditions \citep[e.g.][]{Stencel_1990}. In this case RT~Vir would have to have $\mathrm{M \geq 4M_\odot}$.
    \end{itemize}
\end{enumerate}

\subsection{Future Work}
\label{sec:futurework}

 The interaction of the ISM with the astrosphere of RT~Vir may act as a secondary source of energy contributing to FIR emission \citep[see e.g.][]{1993ApJ...409..725Y,Ueta_2006}. While outside the scope of this investigation, including an external factor like bow-shock interactions with dust and gas may enhance the accuracy of RT modeling.

Referring to Figure \ref{fig:schem}, the nominal gap between the inner and outer dust shells in our model is $\sim 1'$. Confirming this gap observationally would require interferometric measurements with an angular resolution of 0.2$''$ or better. Instruments like MIRC-X on the CHARA Interferometer (Mt Wilson) could observe the dust shell in the near-infrared (NIR) with sufficient angular resolution to resolve such a dust gap. Similar observations at longer wavelengths (mid-IR) could be obtained with MATISSE on VLTI for southern hemisphere objects. Meanwhile, ALMA could trace the molecular gas emission.

\begin{acknowledgments}

We would like to acknowledge the support of the University of Texas at San Antonio. This research was funded by the National Science Foundation via NSF AST 2106926. B. A. Sargent would like to acknowledge funding via NASA APRA grant 80NSSC21K1468. We thank Dr. Kevin Volk for his M-star template used in our DUSTY models, and Dr. Joe Nuth for his wise counsel and expert advice as we revised the manuscript.
We used the resources of Simbad: Strasbourg astronomical Data Center: \url{https://simbad.u-strasbg.fr} and NASA Infrared Science Archive (IRSA): \url{https://irsa.ipac.caltech.edu/cgi-bin/Radar/nph-discovery} in producing this research.
Data used in this paper from IRSA can be accessed via \dataset[https://doi.org/10.26131/IRSA1]{https://doi.org/10.26131/IRSA1} (AllWISE)
\dataset[https://doi.org/10.26131/IRSA2]{https://doi.org/10.26131/IRSA2} (2MASS);
\dataset[https://doi.org/10.26131/IRSA4]{https://doi.org/10.26131/IRSA4} (IRAS)
\dataset[https://doi.org/10.26131/IRSA74]{https://doi.org/10.26131/IRSA74} (HERSCHEL, MESS);
\dataset[https://doi.org/10.26131/IRSA181]{https://doi.org/10.26131/IRSA181}; (AKARI)
\dataset[https://doi.org/10.26131/IRSA524]{https://doi.org/10.26131/IRSA524} (unWISE);
\dataset[https://doi.org/10.26131/IRSA544]{https://doi.org/10.26131/IRSA544} (GAIA DR3).
GALEX data used in this paper were obtained from the Mikulski Archive for Space Telescopes (MAST) at the Space Telescope Science Institute. The specific observations analyzed can be accessed via \dataset[https://doi.org/10.17909/T9H59D]{https://doi.org/10.17909/T9H59D}.
Finally, we would like to acknowledge that this investigation has been carried out on lands of the Jumanos, Coahuiltecan, Lipan Apache, and Tonkawa peoples.

\end{acknowledgments}

\vspace{5mm}

\software{DUSTY \citep{Nenkova_1999},  \\
          ProbabilityShapeDistributions (\href{https://github.com/astroseandillon/AluminumOxide/blob/main/Code/probability_shape_distributions.py}{github.com/ \\
          astroseandillon/AluminumOxide}), 
          }

\clearpage

\appendix

\section{Photometry Table}

\centering
\begin{longtable}{l@{\hspace{0mm}}c@{\hspace{0mm}}p{4.2cm}@{\hspace{0mm}}c@{\hspace{0mm}}c}
    \caption{Photometry sources used in Figures \ref{fig:SED} and \ref{fig:SEDModel}} 
    \label{tab:fullPhot} \\

    \hline 
    \multicolumn{1}{c}{\textbf{$\lambda(\mu m)$} } & 
    \textbf{Flux (Wm$^{-2}$)} & \multicolumn{1}{c}{\textbf{Catalog}} & 
    \textbf{Source} &
    \textbf{Observatory} \\ \hline 
    \endfirsthead

    \multicolumn{3}{c}{series \tablename\ \thetable{} continued} \\
    \hline 
    \multicolumn{1}{c}{\textbf{$\lambda(\mu m)$}} & \textbf{Flux (Wm$^{-2}$)} & \multicolumn{1}{c}{\textbf{Catalog}} & 
    \textbf{Source} &
    \textbf{Observatory}\\ \hline 
    \endhead

    \hline
    \endfoot

    \hline \hline
    \endlastfoot

        0.1529 & 1.19E-16 & J/ApJ/841/33/table1 & \cite{Montez_2017} & GALEX   \\ 
        0.1529 & 2.07E-16 & J/ApJ/841/33/table1 & \cite{Montez_2017} & GALEX   \\ 
        0.2312 & 3.14E-16 & I/353/gsc242 & \cite{Lasker_2008} & GALEX   \\ 
        0.2312 & 2.40E-16 & J/ApJ/841/33/table1 & \cite{Montez_2017} & GALEX   \\ 
        0.2312 & 7.58E-16 & J/ApJ/841/33/table1 & \cite{Montez_2017} & GALEX   \\ 
        0.3498 & 8.19E-13 & II/379/smssdr4 & \cite{Onken_2024} & Sky Mapper \\ 
        0.3498 & 8.77E-13 & II/379/smssdr4 & \cite{Onken_2024} & Sky Mapper \\ 
        0.3519 & 2.43E-12 & I/353/gsc242 & \cite{Lasker_2008} & SDSS \\ 
        0.3519 & 2.41E-12 & V/154/sdss16 & \cite{Ahumada_2020} & SDSS \\ 
        0.3519 & 4.01E-12 & V/154/sdss16 & \cite{Ahumada_2020} & SDSS \\ 
        0.3871 & 2.50E-12 & II/379/smssdr4 & \cite{Onken_2024} & Sky Mapper \\ 
        0.3871 & 2.54E-12 & II/379/smssdr4 & \cite{Onken_2024} & Sky Mapper \\ 
        0.4020 & 1.37E-09 & I/358/varisum & \cite{Gaia_2022} & Gaia DR3 \\ 
        0.4203 & 2.69E-12 & I/239/hip\_main & \cite{Hipparcos_1997} & Hipparcos \\ 
        0.4203 & 2.49E-12 & I/275/ac2002 & \cite{Urban_1998} & Hipparcos \\ 
        0.4442 & 2.54E-12 & I/305/out & \cite{Lasker_2008} & DSS \\ 
        0.4442 & 4.00E-12 & I/305/out & \cite{Lasker_2008} & DSS \\ 
        0.4442 & 3.92E-12 & V/137D/XHIP & \cite{Anderson_2012} & Hipparcos \\ 
        0.4442 & 2.79E-12 & II/336/apass9 & \cite{Henden_2015} & APASS \\ 
        0.4442 & 2.20E-12 & I/342/f3 & \cite{Andruk_2016} & UCAC \\ 
        0.4442 & 3.27E-12 & J/MNRAS/463/4210\newline/ucac4rpm & \cite{Nascimbeni_2016} & UCAC \\ 
        0.4442 & 3.78E-12 & IV/38/tic & \cite{Stassun_2019} & TESS \\ 
        0.4772 & 3.15E-12 & II/349/ps1 & \cite{Chambers_2016} & Pan-STARRS \\ 
        0.4772 & 8.76E-12 & II/349/ps1 & \cite{Chambers_2016} & Pan-STARRS \\ 
        0.4820 & 3.12E-12 & I/353/gsc242 & \cite{Lasker_2008} & SDSS \\ 
        0.4820 & 2.04E-12 & II/336/apass9 & \cite{Henden_2015} & SDSS \\ 
        0.4820 & 3.61E-12 & V/154/sdss16 & \cite{Ahumada_2020} & SDSS \\ 
        0.4820 & 3.62E-12 & V/154/sdss16 & \cite{Ahumada_2020} & SDSS \\ 
        0.4968 & 1.02E-11 & II/379/smssdr4 & \cite{Onken_2024} & Sky Mapper \\ 
        0.4968 & 1.13E-11 & II/379/smssdr4 & \cite{Onken_2024} & Sky Mapper \\ 
        0.5036 & 1.15E-11 & I/350/gaiaedr3 & \cite{Gaia_2021} & Gaia \\ 
        0.5319 & 8.72E-12 & I/239/hip\_main & \cite{Hipparcos_1997} & Hipparcos \\ 
        0.5470 & 1.01E-11 & I/360/syntphot & \cite{Gaia_2022} & Gaia \\ 
        0.5537 & 2.71E-11 & II/122B/merged & \cite{Mermilliod_1987} & Various \\ 
        0.5537 & 9.33E-12 & I/239/hip\_main & \cite{Hipparcos_1997} & Hipparcos \\ 
        0.5537 & 1.48E-11 & I/338/kmac3 & \cite{Karbovsky_2016} & AOTS \\ 
        0.5537 & 9.20E-12 & J/MNRAS/463/4210\newline/ucac4rpm & \cite{Nascimbeni_2016} & UCAC \\ 
        0.5537 & 7.41E-12 & II/366/catalog & \cite{Jayasinghe_2018} & ASAS-SN \\ 
        0.5537 & 9.61E-12 & IV/38/tic & \cite{Stassun_2019} & TESS \\ 
        0.5822 & 1.94E-10 & I/350/gaiaedr3 & \cite{Gaia_2021} & Gaia \\ 
        0.6041 & 2.79E-11 & II/379/smssdr4 & \cite{Onken_2024} & Sky Mapper \\ 
        0.6041 & 5.96E-11 & II/379/smssdr4 & \cite{Onken_2024} & Sky Mapper \\ 
        0.6126 & 2.58E-11 & II/349/ps1 & \cite{Chambers_2016} & Pan-STARRS \\ 
        0.6126 & 6.64E-11 & II/349/ps1 & \cite{Chambers_2016} & Pan-STARRS \\ 
        0.6226 & 1.72E-11 & I/345/gaia2 & \cite{Gaia_2018} & Gaia DR2 \\ 
        0.6247 & 2.54E-11 & I/353/gsc242 & \cite{Lasker_2008} & SDSS \\ 
        0.6247 & 1.90E-11 & V/154/sdss16 & \cite{Ahumada_2020} & SDSS \\ 
        0.6247 & 1.92E-11 & V/154/sdss16 & \cite{Ahumada_2020} & SDSS \\ 
        0.6247 & 2.80E-11 & I/360/syntphot & \cite{Gaia_2022} & SDSS \\ 
        0.6469 & 7.11E-11 & I/360/syntphot & \cite{Gaia_2022} & Gaia \\ 
        0.6730 & 1.46E-11 & I/353/gsc242 & \cite{Lasker_2008} & Gaia \\ 
        0.6730 & 1.46E-11 & J/ApJ/867/105/refcat2 & \cite{Tonry_2018} & Gaia \\ 
        0.6730 & 1.43E-11 & J/A+A/623/A72/hipgpma & \cite{Kervella_2019} & Gaia \\ 
        0.7480 & 3.57E-10 & II/349/ps1 & \cite{Chambers_2016} & Pan-STARRS \\ 
        0.7621 & 4.97E-10 & I/350/gaiaedr3 & \cite{Gaia_2021} & Gaia \\ 
        0.7635 & 4.62E-19 & I/353/gsc242 & \cite{Lasker_2008} & SDSS \\ 
        0.7635 & 1.17E-12 & I/353/gsc242 & \cite{Lasker_2008} & SDSS \\ 
        0.7635 & 4.61E-19 & V/154/sdss16 & \cite{Ahumada_2020} & SDSS \\ 
        0.7635 & 1.92E-12 & V/154/sdss16 & \cite{Ahumada_2020} & SDSS \\ 
        0.7635 & 1.25E-09 & I/360/syntphot & \cite{Gaia_2022} & SDSS \\ 
        0.7713 & 4.89E-11 & II/379/smssdr4 & \cite{Onken_2024} & Sky Mapper \\ 
        0.7713 & 8.62E-11 & II/379/smssdr4 & \cite{Onken_2024} & Sky Mapper \\ 
        0.7713 & 1.58E-10 & II/379/smssdr4 & \cite{Onken_2024} & Sky Mapper \\ 
        0.7713 & 1.87E-10 & II/379/smssdr4 & \cite{Onken_2024} & Sky Mapper \\ 
        0.8652 & 9.84E-10 & II/349/ps1 & \cite{Chambers_2016} & Pan-STARRS \\ 
        0.9018 & 1.16E-11 & I/353/gsc242 & \cite{Lasker_2008} & SDSS \\ 
        0.9018 & 1.65E-11 & V/154/sdss16 & \cite{Ahumada_2020} & SDSS \\ 
        0.9596 & 1.61E-09 & II/349/ps1 & \cite{Chambers_2016} & Pan-STARRS \\ 
        1.2390 & 3.47E-09 & I/280B/ascc & \cite{Kharchenko_2001} & 2MASS \\ 
        1.2500 & 3.53E-09 & II/246/out & \cite{Cutri_2003} & 2MASS \\ 
        1.2632 & 3.64E-09 & J/ApJS/190/203/var & \cite{Price_2010} & COBE \\ 
        1.6300 & 4.36E-09 & II/246/out & \cite{Cutri_2003} & 2MASS \\ 
        1.6495 & 4.35E-09 & I/280B/ascc & \cite{Kharchenko_2001} & 2MASS \\ 
        2.1638 & 3.14E-09 & I/280B/ascc & \cite{Kharchenko_2001} & 2MASS \\ 
        2.1900 & 3.00E-09 & II/246/out & \cite{Cutri_2003} & 2MASS \\ 
        2.2213 & 3.03E-09 & J/ApJS/190/203/var & \cite{Price_2010} & COBE \\ 
        2.7001 & 2.13E-09 & II/161/catalog & \cite{Sweeny_1995} & TMSS \\ 
        3.3500 & 1.44E-10 & I/353/gsc242 & \cite{Lasker_2008} & WISE \\ 
        3.3500 & 5.08E-11 & II/311/wise & \cite{Cutri_2012} & WISE \\ 
        3.3500 & 5.06E-11 & II/338/catalog & \cite{Abrahamyan_2015} & WISE \\ 
        3.3500 & 1.10E-09 & II/363/unwise & \cite{Schlafly_2019} & WISE \\ 
        3.5221 & 1.18E-09 & J/ApJS/190/203/var & \cite{Price_2010} & COBE \\ 
        4.6000 & 5.68E-11 & I/353/gsc242 & \cite{Lasker_2008} & WISE \\ 
        4.6000 & 6.36E-11 & II/311/wise & \cite{Cutri_2012} & WISE \\ 
        4.6000 & 6.40E-11 & II/338/catalog & \cite{Abrahamyan_2015} & WISE \\ 
        4.6000 & 7.35E-10 & II/363/unwise & \cite{Schlafly_2019} & WISE \\ 
        4.8896 & 4.73E-10 & J/ApJS/190/203/var & \cite{Price_2010} & COBE \\ 
        8.6100 & 1.86E-10 & II/297/irc & \cite{Ishihara_2010} & AKARI \\ 
        11.559 & 5.95E-11 & I/353/gsc242 & \cite{Lasker_2008} & WISE \\ 
        11.559 & 8.88E-11 & II/311/wise & \cite{Cutri_2012} & WISE \\ 
        11.590 & 1.43E-10 & II/156A/main & \cite{Moshir_1990} & IRAS \\ 
        11.590 & 1.51E-10 & I/270/cpirss01 & \cite{Hindsley_1994} & IRAS \\ 
        11.590 & 2.83E-13 & V/98/msx & \cite{Egan_1996} & IRAS \\ 
        11.590 & 1.03E-10 & J/MNRAS/471/770/table2 & \cite{McDonald_2017} & IRAS \\ 
        18.389 & 5.29E-11 & II/297/irc & \cite{Ishihara_2010} & AKARI \\ 
        22.090 & 2.18E-11 & I/353/gsc242 & \cite{Lasker_2008} & WISE \\ 
        22.090 & 3.04E-11 & II/311/wise & \cite{Cutri_2012} & WISE \\ 
        23.880 & 3.44E-11 & II/156A/main & \cite{Moshir_1990} & IRAS \\ 
        23.880 & 3.60E-11 & I/270/cpirss01 & \cite{Hindsley_1994} & IRAS \\ 
        23.880 & 2.56E-11 & J/MNRAS/471/770/table2 & \cite{McDonald_2017} & IRAS \\ 
        61.849 & 2.19E-12 & II/156A/main & \cite{Moshir_1990} & IRAS \\ 
        61.849 & 2.40E-12 & J/A+AS/93/121/table4 & \cite{Nyman_1992} & IRAS \\ 
        61.849 & 2.41E-12 & I/270/cpirss01 & \cite{Hindsley_1994} & IRAS \\ 
        101.94 & 5.55E-13 & II/125/main & \cite{IRAS_1988} & IRAS \\ 
        101.94 & 5.89E-13 & II/156A/main & \cite{Moshir_1990} & IRAS \\ 
        101.94 & 5.59E-13 & J/A+A/629/A94/table1 & \cite{Diaz-Luis_2019} & IRAS \\ \hline

\end{longtable}

\bibliography{P2024.bib,drift_vel.bib,MolecularFeatures.bib,Appendix.bib,Perrin1998}{}
\bibliographystyle{aasjournal}

\end{document}